\documentclass[twocolumn]{autart}

\usepackage{graphicx}
\usepackage[compress]{natbib}
\usepackage{enumerate}
\usepackage{mathtools,amsmath,amssymb}
\usepackage{etoolbox}
\AtEndEnvironment{pf}{\null\hfill$\square$}%
\usepackage{xcolor}

\usepackage{aligned-overset}

\usepackage{tikz,pgfplots}\pgfplotsset{compat=newest}
\usepackage{algorithm,algpseudocode}
\algdef{SE}[DOWHILE]{Do}{doWhile}{\algorithmicdo}[1]{\algorithmicwhile\ #1}%

\usepackage{multicol}
\usepackage{booktabs}
\usepackage{setspace}

\usepackage{abbreviations}
\usepackage[bb=dsserif]{mathalpha} 

\renewcommand{\pmod}[1]{\ (\mathrm{mod}\ #1)}

\renewenvironment{thebibliography}[1]{%
    \begin{oldthebibliography}{#1}%
    \setlength{\parskip}{0ex}%
    \setlength{\itemsep}{0ex}%
}%
{%
    \end{oldthebibliography}%
}
\begin{document}
\setlength{\abovedisplayshortskip}{0.6ex plus1ex minus1ex}
\setlength{\abovedisplayskip}{0.6ex plus1ex minus1ex}
\setlength{\belowdisplayshortskip}{0.9ex plus1ex minus1ex}
\setlength{\belowdisplayskip}{0.9ex plus1ex minus1ex}
\begin{frontmatter}

\title{Decrypting Nonlinearity: Koopman Interpretation and Analysis of Cryptosystems\thanksref{footnoteinfo}}

\thanks[footnoteinfo]{
    F.\ Allgöwer is thankful that this work was funded by the Deutsche Forschungsgemeinschaft (DFG, German Research Foundation) under Germany's Excellence Strategy -- EXC 2075 -- 390740016 and within grant AL 316/13-2 -- 285825138 and AL 316/15-1 -- 468094890. 
    R.\ Strässer and S.\ Schlor thank the Graduate Academy of the SC SimTech for its support.
}

\author{Robin Str\"asser}, 
\author{Sebastian Schlor},  
\author{Frank Allg\"ower}
\address{
    University of Stuttgart, Institute for Systems Theory and Automatic Control, 70550 Stuttgart, Germany\\
    (e-mail: \{straesser,schlor,allgower\}@ist.uni-stuttgart.de)
}

\begin{abstract}                
    Public-key cryptosystems rely on computationally difficult problems for security, traditionally analyzed using number theory methods. 
    In this paper, we introduce a novel perspective on cryptosystems by viewing the Diffie-Hellman key exchange and the Rivest-Shamir-Adleman cryptosystem as nonlinear dynamical systems. 
    By applying Koopman theory, we transform these dynamical systems into higher-dimensional spaces and analytically derive equivalent purely linear systems.
    This formulation allows us to reconstruct the secret integers of the cryptosystems through straightforward manipulations, leveraging the tools available for linear systems analysis.
    Additionally, we establish an upper bound on the minimum lifting dimension required to achieve perfect accuracy.
    Our results on the required lifting dimension are in line with the intractability of brute-force attacks.
    To showcase the potential of our approach, we establish connections between our findings and existing results on algorithmic complexity.
    Furthermore, we extend this methodology to a data-driven context, where the Koopman representation is learned from data samples of the cryptosystems.
    \vspace*{-1.25\baselineskip}
\end{abstract}


\end{frontmatter}

\section{Introduction}\label{sec:introduction}
\vspace*{-0.5\baselineskip}
In the long history of cryptography, new concepts have been proposed and analyzed. While some have been found to be secure and effective, others have been broken and abandoned. One major innovation was the introduction of public-key cryptography, which relies on \emph{hard} problems like integer factorization or discrete logarithms for its security.
Two of the oldest and still widely used public-key protocols are the \emph{Diffie-Hellman} (DH) key exchange and the \emph{Rivest–Shamir–Adle\-man} (RSA) cryptosystem.
In DH, the generation of a shared secret over an insecure communication channel is achieved through the operation
$ 
    c = m^e \pmod p 
$
with a ciphertext $c$, a public modulus $p$, a public base $m$, and a secret integer $e$, whereas for sending encrypted messages with RSA the same operation with a public exponent $e$ and a secret message $m$ is used.
The security of the schemes is traditionally studied from a number theory or computational complexity perspective.
Alternatively, we propose an approach to interpret the nonlinear mapping from the plaintext to the ciphertext as the effect of a dynamical system with an unknown parameter, either $e$ or $m$~\citep{schmitz:2008}.
\vspace*{-0.625\baselineskip}

\emph{Related work:~}
Dynamical systems have been of interest to cryptologists for a long time.
Especially chaotic dynamical systems have been used to propose new cryptosystems, as the characteristic properties of chaotic systems, namely sensitive dependence on initial conditions and topological mixing, are desired~\citep[cf.][]{shannon:1949,schmitz:2001}.
To this end, several chaos based cryptosystems were introduced, e.g., in~\cite{pejas:skrobek:2010,habutsu:nishio:sasase:mori:1991}.
However, most of them were proven slow or even insecure~\citep{biham:1991,holt:2017}.
Consequently, several design rules for chaos-based cryptosystems were introduced~\citep{alvarez:li:2006,kelber:schwarz:2005,schmitz:2001}.
The relation between traditional cryptography and cryptosystems utilizing chaotic dynamical systems was explored in previous studies~\citep{millerioux:amigo:daafouz:2008,kocarev:2001}.
A particularly popular approach is the usage of cellular automata as the basic source of deterministic diffusion for the cryptosystem~\citep{martindelrey:mateus:sanchez:2005,hernandez:martindelrey:hernandez:2002,gutowitz:1993}.
Moreover, the analysis of discrete dynamics over finite fields was investigated, e.g., in~\cite{park:2009,colon-reyes:jarrah:laubenbacher:sturmfels:2006,elspas:1959}, and the complexity of trajectories was analyzed in~\cite{brudno:1978,batterman:white:1996}.
\\
A different viewpoint was taken by~\cite{schmitz:2008}, where instead of creating a cryptosystem based on dynamical systems, RSA and other cryptosystems were interpreted as dynamical systems.
Guessing the secret key was translated into guessing the value of a parameter of the dynamical system.
However, the presence of the nonlinear modulo operation poses a significant challenge in system analysis.
This nonlinearity is crucial for maintaining the security of the cryptosystems, as computing the discrete logarithm modulo $p$ is generally an intractable problem~\citep{shoup:1997,maurer:1994}.
\\
As a remedy, we make use of Koopman operator theory for the analysis of the dynamical system, which gained increasing attention in recent years~\citep{koopman:1931,koopman:neumann:1932}.
Following Koopman's seminal work, subsequent studies by~\cite{mezic:banaszuk:2004,mezic:2005} revisited the Koopman operator and proposed its application for prediction and control. 
Since then, numerous contributions have expanded upon this foundation.
These include investigations into global stability properties~\citep{mauroy:mezic:2016}, estimation~\citep{netto:mili:2018}, spectral analysis~\citep{korda:putinar:mezic:2020}, extensions to controlled systems~\citep{korda:mezic:2018a,strasser:berberich:allgower:2023}, and the development of numerical methods for data-driven approximation of the Koopman operator, e.g., Extended Dynamic Mode Decomposition (EDMD)~\citep{williams:kevrekidis:rowley:2015}, along with studies on its properties~\citep{korda:mezic:2018b,haseli:cortes:2022}.
Further, the Koopman operator was proven useful in various applications, including fluid dynamics~\citep{arbabi:mezic:2017a}, power grids~\citep{korda:susuki:mezic:2018},
and robotics~\citep{bruder:remy:vasudevan:2019}. These applications encompass systems with stable attractors as well as limit cycles, i.e., periodic orbits.
Moreover, Koopman operator theory has shown interesting results in the field of number theory, e.g., for the Collatz conjecture~\citep{lagarias:2011,leventides:poulios:2021}.
Related to cryptography,~\cite{byerly:drager:lance:lee:2003} studied the lifting of stream ciphers to higher dimensions in a complex space. Further,~\cite{sule:2022,sule:2021,anantharaman:sule:2021} proposed a Koopman representation for the analysis of nonlinear systems over finite fields using local inversion of maps and observability theory, where iterates over the encryption map describe a periodic dynamical system.
On the other hand,~\cite{schlor:strasser:allgower:2023} exploit Koopman operator theory in combination with a dynamical system corresponding to the encryption map itself to analyze the cryptosystem.

\emph{Contribution:~}
While existing studies on cryptosystems mainly rely on number theory, we offer a different perspective by examining cryptosystems through the lens of dynamical systems theory.
In particular, we build on the analysis in~\cite{schlor:strasser:allgower:2023} for the DH key exchange to generalize the therein derived results to the RSA cryptosystem.
While this is the first work where a linear dynamical representation is derived for RSA, we also provide additional illustrations and explanations for the linear representation of DH.
Here, the ciphertext is interpreted as the endpoint of a trajectory of a nonlinear dynamical system where the plaintext is the length of the trajectory or a parameter of the system, respectively.
Although there exist numerous tools for analyzing nonlinear systems, the theory for linear systems is more comprehensive and provides deeper insights.
Therefore, we utilize Koopman theory to transform the nonlinear system into an equivalent linear system, enabling the reconstruction of the secret integers.
In particular, we can view the cryptosystem through the lens of systems theory and re-establish classical results on the complexity of attacks on the cryptosystems.
More precisely, this reinterpretation offers a valuable insight: Breaking the encryption involves estimating either the trajectory's length or the unknown parameter of the dynamical system.
Additionally, we analytically derive an upper bound on the minimum lifting dimension required to achieve an exact linear representation. 
Our findings support the assumption on the underlying \emph{hard} problems, as the dimension of the linear system cannot be significantly reduced, mirroring the difficulty of brute-force attacks.
We compare these results to a classical metric from algorithmic complexity theory.
Further, we extend our approach to a purely data-driven context, where we learn the Koopman representation of the cryptosystem from data.
While our results do not lead to new methods breaking existing cryptosystems, we offer a new perspective on classical cryptographic algorithms.
The results can build a bridge between cryptography and systems theory for further fruitful exchange of ideas.

\emph{Outline:~}
After providing the preliminaries for our approach in Section~\ref{sec:prelim} we formally introduce the considered cryptosystems and their dynamical counterparts in Section~\ref{sec:cryptosystems}.
In Section~\ref{sec:reconstruction-of-secrets}, we show how the secrets of the cryptosystems can be reconstructed based on a linear system representation, which we derive in Section~\ref{sec:Koopman}.
We devote Section~\ref{sec:minimal-observable-number} to the derivation of the bounds on the required lifting dimensions for the linear representation. 
Section~\ref{sec:data-Koopman} provides a purely data-based treatment of the problem, before Section~\ref{sec:summary-outlook} concludes the paper.

\textbf{Notation:}
We denote the natural numbers excluding zero and the integers by $\bbN$ and $\bbZ$, respectively.
Additionally, $\bbN_0=\bbN \cup\{0\}$ refers to the set of nonnegative integers including zero, and $\bbN_{a:b}$ denotes all nonnegative integers in the closed interval $[a,b]$.
Further, for vectors $a,b\in\bbR^n$ we define $a \oslash b\in\bbR^n$ as the entry-wise division of the corresponding elements, i.e., $(a \oslash b)_j = a_j/b_j$.  
We also make use of the (complex) natural logarithm, denoted by $\ln(z)$, where $z\in\bbC$ is a complex number. 
When $z$ is expressed in polar form $z=r\exp(i\theta)$, where $r>0$ and $\theta\in\bbR$ are the magnitude and angle (in radians) of $z$, respectively, all complex logarithms of $z$ take the form $\ln(r) + i(\theta + 2\pi\ell)$ for any $\ell\in\bbZ$ since 
\begin{equation}\label{eq:notation:comlex-log-periodicity}
    \exp({i\theta}) = \exp({i\theta} + i2\pi \ell).
\end{equation}
An integer $m$ is called a \emph{quadratic residue} modulo $p$ (abbreviated as $m\,\mathrm{R}\,p$) if there exists an integer $x$ such that $
    m=x^2 \pmod{p}
$.
Two integers $p_1,p_2$ are \emph{relatively prime} (also called \emph{coprime}) if their greatest common divisor is equal to 1.
The set of integers from $\bbN_{1:p}$ that are relatively prime to $p$ forms the multiplicative group $\bbZ_p^*$ of integers modulo $p$.
Additionally, we abbreviate $\gamma_1 \pmod{p} = \gamma_2 \pmod{p}$ by $\gamma_1 \overset{{\pmod{p}}}{=}\gamma_2$.
A number $m$ is called \emph{primitive root modulo $p$} if, for any integer $c$ coprime to $p$, there exists an exponent $k\in\bbN$ such that $m^k \overset{{\pmod{p}}}{=} c$. 
Alternatively, $m$ is referred to as a \emph{generator} of the multiplicative group $\bbZ_p^*$ or a generator of $p$.
By $\varPhi(p)$ we denote Euler's totient function which determines the order of $\bbZ_p^*$,
i.e., the number of integers $j\in\bbN_{1:p}$ for which its greatest common divisor is equal to 1. For any prime number $\rho$ holds $\varPhi(\rho) = \rho - 1$. If two numbers $\rho_1$ and $\rho_2$ are relatively prime, then $\varPhi(\rho_1\rho_2) = \varPhi(\rho_1)\varPhi(\rho_2)$.
By $\lambda(p)$ we denote the Carmichael function~\citep[cf.][]{carmichael:1910} of a positive integer $p$ which returns the smallest positive integer $\ell$ such that $m^\ell \pmod{p} = 1$ for every integer $m$ coprime to $p$. 
Moreover, $\lambda(p)$ is the exponent of the multiplicative group $\bbZ_p^*$. 
For any prime number $\rho$ holds $\lambda(\rho) = \varPhi(\rho)$. 
If two numbers $\rho_1$, $\rho_2$ are relatively prime, then $\lambda(\rho_1\rho_2) = \mathrm{lcm}(\lambda(\rho_1),\lambda(\rho_2))$, where $\mathrm{lcm}(a,b)$ denotes the least common multiple of $a,b\in\bbN$.
Further, we write $\mathbb{1}$ for a vector containing only ones.

\vspace*{-0.7\baselineskip}
\section{Preliminaries}\label{sec:prelim}
\vspace*{-0.7\baselineskip}
In this section, we review the theoretical findings that serve as the foundation for our approach in the remainder of the paper.

\vspace*{-0.7\baselineskip}
\subsection{Koopman operator}\label{sec:prelim-Koopman}
\vspace*{-0.7\baselineskip}
Consider a nonlinear dynamical system represented by 
\begin{equation}\label{eq:nonlinear-Koopman}
    x_{k+1} = f(x_k).
\end{equation}
Here, $x_k\in\cX$ represents the state of the system at time step $k\in\bbN_0$ and $f: \cX \to \cX$ is a nonlinear state transition map defined on the state space $\cX\subseteq\bbR^n$.
To analyze the evolution of this system, we introduce the Koopman operator $\cK$, which provides a different perspective by viewing the system's behavior through scalar functions called \emph{observables}. 
The Koopman operator is defined by
\begin{equation}\label{eq:Koopman-equation}
    \cK h_0 = h_0 \circ f,
\end{equation}
where $h_0:\cX \to \bbC$ represents an observable, $\cF$ denotes a space of functions that are invariant under the action of the Koopman operator, and the symbol $\circ$ represents function composition.
For any given observable $h_0$ and state $x_k$,~\eqref{eq:Koopman-equation} is equivalent to $\cK h_0(x_k) = h_0(x_{k+1})$, indicating that the Koopman operator applied to an observable at state $x_k$ yields the value of that observable at the next state $x_{k+1}$.
It is important to note that although the Koopman operator is \emph{linear}, it is typically infinite-dimensional even if the underlying nonlinear dynamical system is finite-dimensional.
Representing a nonlinear dynamical system by an infinite-dimensional linear operator is also related to the Carleman linearization~\citep{carleman:1932}, where the observables are restricted to monomials.
If we select the space of observables $\cF$ such that all components of the state $x_i$, where $i\in\bbN_{1:n}$, are contained in $\cF$, then the Koopman operator fully captures all characteristics of the nonlinear system.
Particularly interesting is the existence of a finite-dimensional set of observables $\cF_n\subseteq \cF$, which remains invariant under the Koopman operator and is rich enough to capture the nonlinear dynamics of the system.
In contrast to a linearization that relies on a first-order Taylor expansion, the linear Koopman operator is able to \emph{globally} describe the behavior of a general nonlinear system.

\vspace*{-0.5\baselineskip}
\subsection{Extended Dynamic Mode Decomposition}\label{sec:EDMD}
\vspace*{-0.5\baselineskip}
To tackle the challenge posed by the infinite-dimensional Koopman operator, we use EDMD for an approximation~\citep{williams:kevrekidis:rowley:2015}. 
This method requires a data trajectory $\{x_k\}_{k=0}^{N}$ of~\eqref{eq:nonlinear-Koopman} and a dictionary of observables $\cD = \{h_j\}_{j=0}^q$, where $h_j\in\cF$. 
Here, $\cF_\cD$ represents the span of the dictionary functions, which are considered as observables. 
The data is organized into two matrices, namely $
X = \begin{bmatrix}
    x_0 & x_1 & \cdots & x_{N-1}
\end{bmatrix}
$ and $
    X_+ = \begin{bmatrix}
        x_1 & x_2 & \cdots & x_{N}
    \end{bmatrix} 
$, and $h : \cX \to \bbC^{q+1}$ defines a vector-valued observable, where $
    h(x) = \begin{bmatrix}
        h_0(x) & h_1(x) & \cdots & h_q(x)
    \end{bmatrix}^\top
$.
The data used for EDMD can come from a single trajectory with $N+1$ data points, $N$ data pairs, or multiple short trajectories.
\\
To simplify the notation, we define $Z=h(X)$, where $
    h(X) = \begin{bmatrix}
        h(x_0) & h(x_1) & \cdots & h(x_{N-1})
    \end{bmatrix}^\top
$ is a matrix formed by applying the observable function $h$ to each data point in $X$. Similarly, $Z_+=h(X_+)$ is obtained by applying $h$ to the corresponding data points in $X_+$. 
Then, the finite-dimensional approximation $K$ of the Koopman operator $\cK$ is obtained through the least-squares optimization problem $\min_K \| Z_+ - K Z \|_\mathrm{F}$, where $\|\cdot\|_\mathrm{F}$ denotes the Frobenius norm. 
The solution is given by $K = Z_+Z^\dagger$ with $Z^\dagger=Z^\top(ZZ^\top)^{-1}$ representing the Moore-Penrose inverse of $Z$.
Under certain assumptions on the space of observables $\cF$ and the choice of the dictionary spanning $\cF_\cD$, the approximation $K$ converges to the true Koopman operator $\cK$ as the number of data points $N+1$ and the dimension of the dictionary $q+1$ tend to infinity~\citep[cf.][]{klus:koltai:schlutte:2016,korda:mezic:2018b}.
In Section~\ref{sec:minimal-observable-number}, we will show that even a finite number of observables $q+1$ results in an exact linear representation of the considered cryptosystems, which are modeled as periodic dynamical systems.

\vspace*{-0.5\baselineskip}
\section{Considered cryptosystems and their dynamical system interpretation}\label{sec:cryptosystems}
\vspace*{-0.5\baselineskip}
After establishing the basic concepts of Koopman theory, we proceed to introduce the DH and RSA cryptosystems for further investigation.

\vspace*{-0.35\baselineskip}
\subsection{The Diffie-Hellman cryptosystem}
\vspace*{-0.35\baselineskip}
The DH key exchange is an early example of a public-key cryptosystem that is still widely used for generating shared secret keys in symmetric cryptography. 
In the original version proposed by~\cite{diffie:hellman:1976}, a large prime number $p$ and a primitive root modulo $p$ denoted as $m$ are publicly shared. The primitive root $m$ is a generator of $\bbZ_p^*$, i.e., the multiplicative group of integers modulo $p$.
In this scheme, two parties each choose a secret number, denoted as $e$ and $d$, respectively, where $e,d\in\bbN_{1:p-1}$. The parties then compute values $c_e$ and $c_d$ as $
    c_e = m^e \pmod p 
$ and $ 
    c_d = m^d \pmod p
$.
These values are made public.
The shared secret key is then obtained by both parties computing 
\begin{alignat*}{2}
    c_{ed} &= c_e^d \pmod p \;&=\;& m^{ed} \pmod p
    ,\\ 
    c_{de} &= c_d^e \pmod p \;&=\;& m^{de} \pmod p
\end{alignat*}
at the respective other party, where $c_{ed} = c_{de}$.
We can express the DH key exchange scheme as a set of dynamical systems. 
The first part of the key exchange, involving the computation of $c_e$ and $c_d$, can be represented by the dynamical system
\begin{equation}\label{eq:dynamical-system}
    x_{k+1} = m x_k \pmod p, \qquad x_0 = 1
\end{equation}
with the endpoints $c_e=x_e$ and $c_d=x_d$ of the trajectories.
Similarly, the second part of the key exchange can be modeled as two separate dynamical systems. For the first party, we have
\begin{equation}\label{eq:sys_c_d}
    y_{k+1} = c_d y_k \pmod p, \qquad y_0 = 1
\end{equation}
leading to $c_{de} = y_e$. For the second party, the system is represented by
\begin{equation}\label{eq:sys_c_e}
    y_{k+1} = c_e y_k \pmod p, \qquad y_0 = 1
\end{equation}
with $c_{ed} = y_d$.
\\
The concept of guessing one of the exponents, e.g., $e$, in the DH scheme can be understood as determining the length of a trajectory of the dynamical system~\eqref{eq:dynamical-system} that connects two known points.
Similarly, guessing the shared secret $c_{ed}$ involves finding the point of the intersection between the trajectories of~\eqref{eq:sys_c_d} and~\eqref{eq:sys_c_e}, which may occur at different time steps $e$ and $d$.
Moreover, the state $x_{ed}$ of~\eqref{eq:dynamical-system} at time step $ed$ matches the intersection point.
With these considerations, two questions arise~\citep{schlor:strasser:allgower:2023}.
\begin{prob}\label{problem:DH_message}
    Can we determine the secret exponents $e$ (and $d$) by analyzing the dynamical system~\eqref{eq:dynamical-system}?
\end{prob}
\begin{prob}\label{problem:DH_key}
    Can the insights gained from the dynamical system's view aid in estimating the shared secret $c_{ed}$?
\end{prob}
These questions aim to explore how the system dynamics relate to the identification of the secret exponents and the estimation of the shared secret in the DH scheme.

\textbf{Cryptographic view~}\emph{
    Problem~\ref{problem:DH_key}, i.e., computing $c_{ed}$ given $c_e$ and $c_d$, is known as the \emph{computational DH problem}~\citep{katz:lindell:2014}.
    The \emph{discrete logarithm problem} in Problem~\ref{problem:DH_message} is at least as hard to solve as the computational DH problem. 
    Currently, there are no efficient classical algorithms known to solve these problems.
}

\vspace*{-0.35\baselineskip}
\subsection{The Rivest–Shamir–Adleman cryptosystem}
\vspace*{-0.35\baselineskip}
RSA is a cryptosystem commonly used for encryption or signatures.
In the following, we describe the simplified RSA scheme~\citep{boneh:1999}.
To set up the encryption scheme, first, two distinct large prime numbers $p_1$ and $p_2$ of the same bit length are needed. The product $p = p_1p_2$ yields the RSA modulus of bit length $\chi$ which is recommended to be at least 1024.
Then, an integer $d$ relative prime to Euler's totient function $\varPhi(p)$ is picked.
As $p_1$ and $p_2$ are prime, we have $\varPhi(p) = (p_1-1)(p_2-1)$.
By the definition of Euler's totient function, $\varPhi(p)$ determines the size of supported messages corresponding to an RSA modulus $p$.
The chosen $d$ serves as the secret key, whereas $e$ as its multiplicative inverse modulo $\varPhi(p)$, i.e., $ed \pmod{\varPhi(p)} = 1$, is the public key.
\\
The encryption works as follows. 
A secret message $m\in \bbZ_p^*$ is encrypted by computing 
$
    c = m^e \pmod p 
$
with the public exponent $e$ and the public modulus $p$ to obtain the ciphertext $c\in\bbN_p$.
The message can be recovered by decrypting with the operation 
$
    m = c^d \pmod p
$.
Analogously to DH, the encryption can be rewritten as the dynamical system~\eqref{eq:dynamical-system} with secret parameter $m$.
The ciphertext is then given as $c=x_e$.
Similarly, the decryption is rewritten as
\begin{equation}\label{eq:RSA_decryptSys}
    y_{k+1} = c y_k \pmod{p}, \qquad y_0 = 1
\end{equation}
and $m=y_d$.
\\
Finding the secret message $m$ without knowing $d$ can now be interpreted as finding the parameter $m$ of the dynamical system based on the first and the $e$-th point of a trajectory~\citep{schmitz:2008}.
The fact that the decryption and encryption are inverse maps can be related to the product $ed$ which is then a period length of the systems' trajectories.
This leads to the following questions.
\begin{prob}\label{problem:RSA_key}
    Can we determine the secret key $d$ by analyzing the dynamical system~\eqref{eq:dynamical-system}?
\end{prob}
\begin{prob}\label{problem:RSA_message}
    Can the insights gained from the dynamical system's view aid in estimating the secret message $m$?
\end{prob}
\textbf{Cryptographic view~}\emph{
    Problem~\ref{problem:RSA_message} is typically referred to as \emph{RSA problem}~\citep{katz:lindell:2014}. Problem~\ref{problem:RSA_key} is known to be equivalent to the factoring problem, which is at least as hard as the RSA problem. There is no classical algorithm known to efficiently break these problems.
}

Due to the common form of the dynamical system representing DH and RSA, we consider system~\eqref{eq:dynamical-system} as a common basis for the analysis in the following.

\vspace*{-0.35\baselineskip}
\section{Reconstruction of secrets}\label{sec:reconstruction-of-secrets}
\vspace*{-0.35\baselineskip}
In order to solve the introduced problems, we propose to use established techniques from linear systems theory. 
Although system~\eqref{eq:dynamical-system} is linear under group operations on $\bbZ_p^*$, the modulo operation is nonlinear when applied to $\bbZ$.
Before constructing a linear representation of the nonlinear system~\eqref{eq:dynamical-system}, we demonstrate how the problems could be addressed if a linear system representation is available.
To this end, we assume the availability of a linear dynamical representation $z_{k+1} = A z_k$ of the underlying cryptosystem~\eqref{eq:dynamical-system}, where $z\in\bbC^N$ denotes the state vector with a given initial state $z_0$ and $A\in\bbC^{N\times N}$ is diagonalizable. 
Then, we can employ well-established tools from linear systems theory to analyze the cryptosystem.
The discussion of Problem~\ref{problem:DH_message} and Problem~\ref{problem:DH_key} in Section~\ref{sec:reconstructE} is due to~\cite[Sec.~4]{schlor:strasser:allgower:2023}. 
This serves as a basis for the discussion of Problem~\ref{problem:RSA_key} and Problem~\ref{problem:RSA_message} in Section~\ref{sec:reconstructD}.
Subsequently, we proceed to derive the linear system using principles from Koopman theory.

\vspace*{-0.35\baselineskip}
\subsection{Reconstruction of the secret integer $e$ (Problem~\ref{problem:DH_message})}\label{sec:reconstructE}
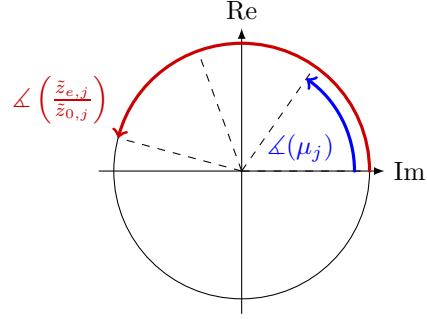
\begin{figure}[bt]
    \centering
    \begin{tikzpicture}
        \def\R{1.7}
        \def\ang{55}
        \def\angTwo{110}
        \def\angThree{165}
        \coordinate (O) at (0,0);
        \coordinate (X) at (\R,0);
        \coordinate (R) at (\ang:\R);
        \draw[-latex] (-1.9,0) -- (1.9,0) node[right]{$\mathrm{Im}$};
        \draw[-latex] (0,-1.9) -- (0,1.9) node[above]{$\mathrm{Re}$};
        \draw[dashed] (O) -- (X) node[midway,below right=0] {};
        \draw[dashed] (O) -- (R) node[midway,right=4,above left=0] {};
        \draw[dashed] (O) -- (\angTwo:\R) node[midway,right=4,above left=0] {};
        \draw[dashed] (O) -- (\angThree:\R) node[midway,right=4,above left=0] {};
        \draw (O) circle (\R);
        \draw[->,black!75,very thick,line cap=round] (X) arc (0:\angThree:\R) node[pos=0.99,above left] {$\measuredangle\left(\frac{\tilde{z}_{e,j}}{\tilde{z}_{0,j}}\right)$};
        \draw[->,black!75,very thick,line cap=round] (0:1.5) arc (0:\ang:1.5) node[pos=0.45,below left] {$\measuredangle(\mu_j)$};
    \end{tikzpicture}
    \caption{Comparison of the angles of the complex numbers occurring in the reconstruction of $e$.}
    \label{fig:complex-angles}
\end{figure}

\vspace*{-0.35\baselineskip}
The representation of the dynamical system in~\eqref{eq:dynamical-system} shows that the ciphertext $c=x_e$ is obtained by evaluating $e$ steps of the system.
If we have a linear representation $z_{k+1} = Az_k$, where $z_k = h(x_k)\in\bbC^{q+1}$ and $z_e = A^e z_0$, we can establish a direct relationship between the secret integer $e$, the ciphertext $c=x_e$, and the initial condition $x_0$. 
The eigendecomposition $A = V \Lambda V^{-1}$, where $\Lambda=\diag(\mu_0,...,\mu_q)$ and $V=\begin{bmatrix}
    v_0 & \cdots & v_q
\end{bmatrix}$ with eigenvectors $v_j$ satisfying $Av_j = \mu_j v_j$, $j\in\bbN_{0:q}$, yields $z_e$ as $z_e = V \Lambda^e V^{-1} z_0$. 
Equivalently, we obtain
\begin{equation}\label{eq:z_e-tilde}
    \tilde{z}_e = \Lambda^e \tilde{z}_0
\end{equation}
by introducing $\tilde{z} = V^{-1} z$. 
We then evaluate each row individually, i.e., $\tilde{z}_{e,j} = \mu_j^e \tilde{z}_{0,j}$ for $j\in\bbN_{0:q}$, such that we obtain the ratio $\frac{\tilde{z}_{e,j}}{\tilde{z}_{0,j}} = \mu_j^e$, where both terms are complex numbers. 
Consequently, we compare their absolute values and respective angles. 
In the following analysis, we focus on the angles and compare them, as depicted in Fig. \ref{fig:complex-angles}.
Given the complex exponential $\mu_j^e$ in the above expression, we rewrite it as
\begin{equation*}
    \mu_j^e
    = \exp(\ln(\mu_j)e)
    \overset{\eqref{eq:notation:comlex-log-periodicity}}{=} \mu_j^e \exp({-i2\pi\ell_j}),
\end{equation*}
where $\ell_j\in\bbZ$. 
Hence, we have
$
    \measuredangle\left(\tfrac{\tilde{z}_{e,j}}{\tilde{z}_{0,j}}\right)
    = e \measuredangle(\mu_j) - 2\pi\ell_j
$
and, thus,
\begin{equation*}
    e = \frac{
        \measuredangle\left(\tfrac{\tilde{z}_{e,j}}{\tilde{z}_{0,j}}\right)
        + 2\pi\ell_j
    }{
        \measuredangle(\mu_j)
    },
\end{equation*}
where $\ell_j$ is chosen such that $e\in\bbN_0$ remains consistent for all $j\in\bbN_{0:q}\setminus\{r\in\bbN_{0:q}| \mu_r\in\bbR\}$.
Note that we exclude all real eigenvalues since they do not provide additional information about the secret integer but only indicate whether it is even or odd.
This aspect will be further addressed in Section~\ref{sec:minimal-observable-number}.
\begin{rem}[Reconstruction of $c_{ed}$ (Problem~\ref{problem:DH_key})]\label{rem:reconstructCed}
    One possible solution to the problem at hand involves solving Problem~\ref{problem:DH_message} to obtain the secret integers~$e$ and~$d$.
    By doing so, the shared secret $c_{ed} = m^{ed} \pmod{p}$ can be easily computed.
    However, there may be a more efficient approach.
    As mentioned earlier, the shared secret can also be determined by intersecting the trajectories of~\eqref{eq:sys_c_d} and~\eqref{eq:sys_c_e}. 
    Suppose we find an intersection point $x_i$ for time steps $e$ and $d$, respectively.
    To confirm that this is the shared secret, we additionally need to verify the condition $x_i = x_{ed} = m^{ed} \pmod{p}$. 
    Therefore, the shared secret corresponds to the first intersection point of the trajectory defined by~\eqref{eq:dynamical-system} with the two trajectories of~\eqref{eq:sys_c_d} and~\eqref{eq:sys_c_e}.
    However, the dynamic interpretation does not provide a straightforward algorithm for finding the intersection. 
    It remains an open question whether there exists an algorithm that is more efficient than a brute-force search along the trajectory values. 
    Further research is required to explore this problem in more detail.
\end{rem}%

\vspace*{-0.35\baselineskip}
\subsection{Reconstruction of the secret key $d$ (Problem~\ref{problem:RSA_key})}\label{sec:reconstructD}
\vspace*{-0.35\baselineskip}
\begin{figure*}[!t]
    \begin{equation}\label{eq:obs-unit-circle-derivation}
        z_{k+1} 
        = \begin{bmatrix}
            \exp\left({i\tfrac{2\pi}{p}(m(m x_k \pmod{p}))}\right) \\
            \vdots \\
            \exp({i\tfrac{2\pi}{p}(m^{q+1}(m x_k \pmod{p}))})
        \end{bmatrix} 
        = \begin{bmatrix}
            \exp\left({i\tfrac{2\pi}{p}(m(m x_k + \ell_0 p))}\right) \\
            \vdots \\
            \exp\left({i\tfrac{2\pi}{p}(m^{q+1}(m x_k + \ell_{q} p))}\right)
        \end{bmatrix}
        = \begin{bmatrix}
            \exp\left({i\tfrac{2\pi}{p}(m^2 x_k)}\right) \\
            \vdots \\
            \exp\left({i\tfrac{2\pi}{p}(m^{q+2} x_k)}\right)
        \end{bmatrix}
    \end{equation}
\medskip
\hrule
\end{figure*}
By the discussion of Section~\ref{sec:cryptosystems}, we can build the dynamical system representation~\eqref{eq:RSA_decryptSys} of the decryption scheme with $y_d=m$ to reconstruct the secret key $d$ based on known $m$ and $c$. 
As we know the public key pair $(p,e)$ in RSA, we can encrypt an arbitrary message $m$ and get the corresponding ciphertext $c$.
Note that the dynamical system representing the RSA decryption is of the same structure as the DH system corresponding to Problem~\ref{problem:DH_message}. Hence, the reconstruction of the secret key $d$ of RSA can be treated analogously to the reconstruction of the secret integer $e$ of the DH cryptosystem in Section~\ref{sec:reconstructE}.
\begin{rem}[Reconstruction of $m$ (Problem~\ref{problem:RSA_message})]\label{rem:reconstructM}
    Analogue to the discussion on Problem~\ref{problem:DH_key} in Remark~\ref{rem:reconstructCed}, we first solve Problem~\ref{problem:RSA_key} to obtain the secret key $d$ with an arbitrary message $\hat{m}$ and then reconstruct the unknown message $m$ corresponding to a given ciphertext $c$ via $m = c^d \pmod{p}$.
    As Problem~\ref{problem:RSA_key} is known to be at least as hard to solve as Problem~\ref{problem:RSA_message}, there might be a more direct and efficient solution to Problem~\ref{problem:RSA_message}.
    A realistic scenario might be that multiple pairs $\{c_i,m_i\}_{i=1}^{\gamma}$ of ciphertext and secret message are available, where the same public key pair $(p,e)$ is used for encryption. 
    Then, we know $c_i = x_{i,e}$, where $x_{i,k} = h_i^{-1}(z_{i,k})$ and $z_{i,e}=A_i^e z_{i,0}$ for all $i\in\bbN_{1:\gamma}$. As each system corresponds to a different message, there is no common linear system matrix $A$, but different matrices $A_i$ depending on $m$. Thus, future work could include the investigation of a linear parameter-varying (LPV) representation of the nonlinear system~\eqref{eq:dynamical-system}. To this end, the LPV parametrization needs to be established first, which may then help to reconstruct the secret message of a ciphertext of interest.
\end{rem}%

\vspace*{-0.35\baselineskip}
\section{Koopman representation}\label{sec:Koopman}
\vspace*{-0.35\baselineskip}
In this section, we establish a linear representation of the dynamical system~\eqref{eq:dynamical-system} using the Koopman operator.
The accuracy and computational efficiency of the lifting process described in Section~\ref{sec:prelim-Koopman} rely on the number and selection of observables. 
Since the lifting is often high-dimensional or even infinite-dimensional, it is important to choose the observables carefully. 
However, determining the optimal dictionary for general nonlinear systems remains an open research question.
In the following, we explore two different options for selecting observable functions for the dynamical system~\eqref{eq:dynamical-system} leading to an exact \emph{finite}-dimensional Koopman representation.

\subsection{Observables on the complex unit circle}\label{sec:observables-complex}
\vspace*{-0.35\baselineskip}
The first considered observables are inspired by the work in~\cite{korda:putinar:mezic:2020} for periodic systems. In particular, we define a $(q+1)$-dimensional lifted state $z_k=h(x_k)=\begin{bmatrix}
    h_0(x_k) & \cdots & h_q(x_k)
\end{bmatrix}^\top$ based on the observables 
\begin{equation*}
    h_j(x) = \exp\left({i\tfrac{2\pi}{p}m^{j+1}x}\right),
\end{equation*}
where $j\in\bbN_{0:q}$ for a given $q$.
\begin{figure}[bt]%
    \centering
    \begin{tikzpicture}
        \def\R{1.7}
        \def\RAng{1.35}
        \def\ang{72}
        \def\angTwo{144}
        \def\angThree{216}
        \def\angFour{288}
        \coordinate (O) at (0,0);
        \draw[-latex] (-1.9,0) -- (1.9,0) node[right]{$\mathrm{Im}$};
        \draw[-latex] (0,-1.9) -- (0,1.9) node[above]{$\mathrm{Re}$};
        \draw[dashed] (O) -- (\ang:\R) node{$\times$} node[above right] {$h_3$};
        \draw[dashed] (O) -- (\angTwo:\R) node{$\times$} node[above left] {$h_2$};
        \draw[dashed] (O) -- (\angThree:\R) node{$\times$} node[below left] {$h_0$};
        \draw[dashed] (O) -- (\angFour:\R) node{$\times$} node[below right] {$h_1$};
        \draw (O) circle (\R);
        \draw[->,black!75,very thick,line cap=round] (0:\RAng) arc (0:\ang:\RAng) node[pos=0.45,below left] {$\frac{2\pi}{p}$};
        \draw[->,black!75,very thick,line cap=round]  (\ang:\RAng) arc (\ang:\angTwo:\RAng) node[pos=0.45,below] {$\frac{2\pi}{p}$};
        \draw[->,black!75,very thick,line cap=round]  (\angTwo:\RAng) arc (\angTwo:\angThree:\RAng) node[pos=0.45,right] {$\frac{2\pi}{p}$};
        \draw[->,black!75,very thick,line cap=round]  (\angThree:\RAng) arc (\angThree:\angFour:\RAng) node[pos=0.45,above] {$\frac{2\pi}{p}$};
        \draw[->,black!75,very thick,line cap=round]  (\angFour:\RAng) arc (\angFour:360:\RAng) node[pos=0.45,above left] {$\frac{2\pi}{p}$};
    \end{tikzpicture}
    \vspace*{-\baselineskip}
    \caption{Observables on complex unit circle exemplary for $p=5$, $m=3$, $x=1$, and $q=3$.}
    \label{fig:observables-complex-unit-circle}
\end{figure}
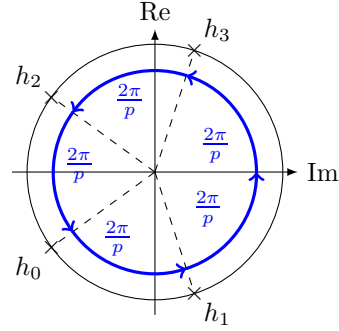%
With these observables, illustrated in Fig.~\ref{fig:observables-complex-unit-circle}, we exploit the periodicity of the modulo operation by dividing the complex unit circle into $p-1$ periodic parts which are equally distributed. 
\begin{rem}
    The chosen observables are particularly interesting because the resulting lifting is related to typical results of the discrete Fourier transform (DFT).
    More precisely, the inverse DFT of a sequence of $p$ complex numbers $\{\cN_j\}_{j=0}^{p-1}$ maps the sequence into another sequence of complex numbers $\{\nu_n\}_{n=0}^{p-1}$, where 
    \begin{equation*}
        \nu_n = \frac{1}{p}\sum_{j=0}^{p-1} \cN_j \exp\left(i\tfrac{2\pi}{p}jn\right).
    \end{equation*}
    In particular, we can express the inverse DFT in terms of the chosen lifting functions $h_j(x)$ and interpret the search for the linear system representation via the (inverse) DFT of two complex sequences. 
    A more detailed investigation of this relation is left for future work.
\end{rem} 
For exact accuracy, the chosen $q$ needs to ensure that the evolution of the lifted state is determined by the linear dynamics $z_{k+1} = A z_k$ for all $k\geq 0$ for a fixed $A$. Investigating $z_{k+1}$, we derive~\eqref{eq:obs-unit-circle-derivation} for some $\ell_0,...,\ell_q\in\bbZ$, where we use $w\pmod{p}=w+\ell p$ for some $\ell\in\bbZ$ in the second equality and the periodicity argument~\eqref{eq:notation:comlex-log-periodicity} of the complex logarithm in the last equality. Note that the first $q$ entries of $z_{k+1}$ are just the last $q$ entries of $z_k$, i.e., 
\begin{equation*}
    z_{k+1} = \begin{bmatrix}
        0 & 1 & \cdots & 0 \\
        \vdots & \vdots & \ddots & \vdots \\
        0 & 0 & \cdots & 1 \\
        0 & 0 & \cdots & 0
    \end{bmatrix}
    z_k + \begin{bmatrix}
        0 \\ \vdots \\ 0 \\ \exp\left({i\tfrac{2\pi}{p}(m^{q+2} x_k)}\right)
    \end{bmatrix}.
\end{equation*}
For a fully linear characterization of the dynamics, we need to find a parameter vector $\alpha\in\bbR^{q+1}$ such that 
\begin{equation*}
    \exp\left({i\tfrac{2\pi}{p}(m^{q+2} x_k)}\right) 
    = \sum_{j=0}^{q} \alpha_j \exp\left({i\tfrac{2\pi}{p}(m^{j+1} x_k)}\right)
\end{equation*}
for all $k\in\bbN_0$. 
In particular, this $\alpha$ needs to be the \emph{same} vector $\alpha$ for all times to obtain a time-invariant system. 
Then, we get $z_{k+1} = A z_k$ with $A$ defined as the companion matrix
\begin{equation}\label{eq:Koopman-A-matrix-RNF}
    A = \begin{bmatrix}
        0 & 1 & \cdots & 0 \\
        \vdots & \vdots & \ddots & \vdots \\
        0 & 0 & \cdots & 1 \\
        \alpha_0 & \alpha_1 & \cdots & \alpha_q
    \end{bmatrix}.
\end{equation}
This vector trivially exists for $q=p-2$ due to the periodicity of the observables on the unit circle, i.e., $h_{p-1}(x)=h_0(x)$ and, thus, $\alpha_0=1$ and $\alpha_j=0$ for $j\in\bbN_{1:p-2}$.
Now, suppose we have established the linear lifted system representation $z_{k+1} = A z_k$ for $z_k = h(x_k)$ and we are interested in the inverse mapping $x_k = h^{-1}(z_k)$. 
For each $j\in\bbN_{0:q}$, we know 
$
    z_{k,j} = \exp\left({i\tfrac{2\pi}{p}m^{j+1}x_k}\right)
$.
Thus, using the complex logarithm we get
$
    -i\tfrac{p}{2\pi}\ln(z_{k,j}) = m^{j+1} x_k - \tilde{\ell}_j p
$
for some $\tilde{\ell}_j\in\bbZ$ or equivalently written in a condensed form 
$
    x_k \mathbb{1} = M^{-1}(\tilde{z}_k + \tilde{\ell} p)
$,
where $\tilde{z}_{k,j} \coloneqq -i\frac{p}{2\pi}\ln(z_{k,j})$, $\tilde{\ell} \coloneqq \begin{bmatrix}
    \tilde{\ell}_0 & \tilde{\ell}_1 & \cdots & \tilde{\ell}_q
\end{bmatrix}^\top$, and $M \coloneqq \diag(m,m^2,...,m^{q+1})$. 
Hence, we have $q+1$ equations to reconstruct $x_k\in\bbN_{1:p-1}$ which all need to be satisfied simultaneously. 
The main idea of the reconstruction is summarized in Algorithm~\ref{alg:reconstruction-xk-complex-unit-circle} and the details are provided in Appendix~\ref{sec:appendix:reconstruction-xk-complex-unit-circle}.
\begin{algorithm}[tb]
    \caption{Reconstruction of $x_k$ based on observables on complex unit circle}\label{alg:reconstruction-xk-complex-unit-circle}
    \textbf{Input:} lifted state $z_k\in\bbC^{q+1}$ \\
    \textbf{Output:} state $x_k\in\bbR$
    \begin{spacing}{0.9}
	    \begin{algorithmic}
	        \State Fix $j\in\bbN_{0:q}$ \Comment{All $j\in\bbN_{0:q}$ lead to the same $x_k$}
	        \State $\tilde{\alpha} \gets -1$
	        \Do 
	            \State $\tilde{\alpha} \gets \tilde{\alpha} + 1$
	            \State $x_\mathrm{guess} \gets \left(\tilde{z}_{k_j} \pmod{p} + \tilde{\alpha} p\right) / \left(m^{j+1} \pmod{p}\right)$
	        \doWhile{$x_\mathrm{guess}\notin\bbN_{1:p-1}$}
	        \State $x_k \gets x_\mathrm{guess}$
	    \end{algorithmic}
    \end{spacing}
\end{algorithm}

\vspace*{-0.35\baselineskip}
\subsection{Observables as a value list}\label{sec:observables-value-list}
\vspace*{-0.35\baselineskip}
In the following, we investigate an alternative observable selection, where we choose the observables simply as a list of subsequent state values.
Here, we define the $(q+1)$-dimensional lifting based on $h_j(x_k) = x_{k+j}$ for each $j\in\bbN_{0:q}$.
Again, the value of $q$ is chosen to ensure that the lifted state follows linear dynamics of the form $z_{k+1} = A z_k$ for all $k\in\bbN_0$, achieving exact accuracy.
Further, the first $q$ entries of $z_{k+1}$ are simply shifts of the entries in $z_k$.
This allows us to represent the lifted dynamics using a linear system with a companion matrix as in~\eqref{eq:Koopman-A-matrix-RNF}, where we seek \emph{one} parameter vector $\alpha\in\bbR^{q+1}$ such that
\begin{equation}\label{eq:closing-condition}
    x_{k+q+1} = \sum_{j=0}^q \alpha_j x_{k+j}
\end{equation}
for all $k\in\bbN_0$. 
Moreover, we can easily retrieve the original state via
$
    x_k = \begin{bmatrix}
        1 & 0 & \cdots & 0
    \end{bmatrix}
    z_k
$.
This structure is closely related to the delay-based lifting functions used in Hankel DMD~\citep{susuki:mezic:2015,arbabi:mezic:2017b}.

The accuracy of the obtained lifting is typically influenced by the number of chosen lifting functions. 
In Section~\ref{sec:minimal-observable-number}, we provide detailed explanations regarding possible selections of $\alpha$ corresponding to different choices of $q$.
Although the subsequent analysis can be derived similarly for the observables presented in Section~\ref{sec:observables-complex} and yield similar results, we restrict ourselves to the observables from this section in the remainder of the paper because they are easier to present and interpret.

\vspace*{-0.35\baselineskip}
\section{Minimal number of observables}\label{sec:minimal-observable-number}
\vspace*{-0.35\baselineskip}
We will now explore the minimum value of $q$ required to obtain a linear representation of the cryptosystems at hand.
To this end, we recall the dynamics~\eqref{eq:dynamical-system} and note that $m$ and $p$ are coprime, as $m$ belongs to $\bbZ_p^*$. 
Consequently, we have $m x_k \neq p$, and the achievable values of $x$ are within the range $\bbN_{1:p-1}$. 
This is due to the modulo operation, which restricts the possible values to at most $p-1$ distinct elements.
Further, we are particularly interested in the period length of the periodic system~\eqref{eq:dynamical-system}, i.e., the number of state evaluations $\tau$ after which we obtain $x_\tau=x_0=1$. To this end, we recall Fermat's little theorem~\citep{hardy:wright:1979}.
\begin{lem}[Fermat's little theorem]\label{lm:Fermat}
    If $p$ is a prime number, then any integer $m<p$ satisfies
    \begin{equation*}
        m^{p-1} \pmod{p} = 1.
    \end{equation*}%
\end{lem}%
\vspace*{-0.65\baselineskip}
The earliest proof of this theorem was presented by Leibniz in an unpublished manuscript around fifty years before the published proof by~\cite{euler:1761}.
Since the modulus $p$ is only a prime number for DH but not for RSA, we recall the following lemma generalizing Fermat's little theorem to any modulus $p\in\bbN$ without restriction to prime numbers.
\begin{lem}[Euler's theorem]\label{lm:Euler}
    Any integer $p\in\bbN$ and any integer $m$ coprime to $p$ satisfy
    \begin{equation*}
        m^{\varPhi(p)} \pmod{p} = 1.
    \end{equation*}%
\end{lem}%
\vspace*{-0.65\baselineskip}
Fermat's little theorem is a special case of Euler's theorem as $\varPhi(p)=p-1$ for any prime number $p$ and any integer $m<p$ that is coprime to a prime number $p$.
As a consequence of Lemma~\ref{lm:Euler}, $x_{\varPhi(p)}$ has again the same value as $x_0=1$ for both DH and RSA, and, thus, a general upper bound on the necessary number of observables is $\varPhi(p)$, i.e., $q=\varPhi(p)-1$.
Thus, as discussed above, all achievable values of the dynamics are contained in the observable vector corresponding to the linear system representation
\begin{equation}\label{eq:dynamics-linear-system-upper-bound}
    \begin{bmatrix}
        x_{k+1} \\ \vdots \\ x_{k+\varPhi(p)-1} \\ x_{k+\varPhi(p)}
    \end{bmatrix} 
    = \begin{bmatrix}
        0 & 1 & \cdots & 0 \\
        \vdots & \vdots & \ddots & \vdots \\
        0 & 0 & \cdots & 1 \\
        1 & 0 & \cdots & 0
    \end{bmatrix}
    \begin{bmatrix}
        x_{k} \\ x_{k+1} \\ \vdots \\ x_{k+\varPhi(p)-1}
    \end{bmatrix}.
\end{equation}

\vspace*{-0.35\baselineskip}
\subsection{Minimal number of observables for DH}\label{sec:minimal-observable-number-DH}
\vspace*{-0.35\baselineskip}
Next, we demonstrate that it is possible to decrease the required number of observables and derive the minimum number of observables for the ones introduced in Section~\ref{sec:observables-value-list}.
To establish the minimum number, we begin by revisiting Euler's criterion~\citep{euler:1763}.
\begin{lem}[Euler's criterion]\label{lm:Euler-criterion}
    If $p$ is an odd prime number, then for any coprime integer $m$ holds%
    \begin{equation}\label{eq:Euler-criterion}%
        m^{\frac{p-1}{2}} \overset{{\pmod{p}}}{=} \begin{cases}
            1 & \text{if}\>m\,\mathrm{R}\,p, \\
            -1 & \text{else}.
        \end{cases}
    \end{equation}%
\end{lem}%
This result is also known as Legendre's symbol~\citep{legendre:1798}.%
\begin{cor}[{\citeauthor{schlor:strasser:allgower:2023},~\citeyear{schlor:strasser:allgower:2023},~Cor.~6}]\label{cor:Euler-criterion-DH}%
    If $p$ is an odd prime number with $p>3$, then for any generator $m$ of the multiplicative group of integers modulo $p$ holds 
    \begin{equation*}
        m^{\frac{p-1}{2}} \overset{{\pmod{p}}}{=} -1.
    \end{equation*}
    Moreover, $m$ is no quadratic residue modulo $p$.
\end{cor}%
Now we can characterize the minimal number of lifting functions to obtain a linear representation of the DH cryptosystem, which was first derived by~\cite{schlor:strasser:allgower:2023}.%
\begin{thm}[{\citeauthor{schlor:strasser:allgower:2023},~\citeyear{schlor:strasser:allgower:2023},~Thm.~7}]\label{thm:minimal-number-observables-DH}%
    Let $p$ be an odd prime number with $p>3$ and $m$ a primitive root modulo $p$ corresponding to a DH cryptosystem. Then, the minimal lifting dimension to obtain a linear representation of the cryptosystem with the observables in Section~\ref{sec:observables-value-list} is $\tilde{q}+1$, where $\tilde{q} = (p-1)/2$.
\end{thm}%
\begin{pf}%
    Despite this theorem was already proven in~\cite[Thm.~7]{schlor:strasser:allgower:2023}, we include the proof for the sake of completeness while providing additional explanations and a more precise argumentation.
    \\
    To establish the theorem, we examine the obtained linear representation by varying the selection of $q$. 
    During this investigation, we explore the implications on the expressiveness of the lifted linear system that results with the observables discussed in Section~\ref{sec:observables-value-list}. 
    To this end, we consider a general natural number $q\in\bbN$ and derive the linear representation 
    \begin{equation}\label{eq:linear-dynamics-general-q}
        \begin{bmatrix}
            x_{k+1} \\ \vdots \\ x_{k+q} \\ x_{k+q+1}
        \end{bmatrix}
        = \begin{bmatrix}
            0 & 1 & \cdots & 0 \\
            \vdots & \vdots & \ddots & \vdots \\
            0 & 0 & \cdots & 1 \\
            \alpha_0 & \alpha_1 & \cdots & \alpha_{q}
        \end{bmatrix}
        \begin{bmatrix}
            x_{k} \\ x_{k+1} \\ \vdots \\ x_{k+q}
        \end{bmatrix}.
    \end{equation}
    The choice of the parameter vector $\alpha$ is crucial for ensuring that~\eqref{eq:closing-condition} holds for all $k\in\bbN_0$. 
    Due to Lemma~\ref{lm:Fermat}, the periodicity of $x_k$ is given by 
    \begin{equation}\label{eq:periodicity-xk}
        x_{k+p-1} = m^{p-1} x_k \pmod{p} = x_k,
    \end{equation}
    such that we only need to focus on $k\in\bbN_{0:p-2}$.
    Specifically, we aim to determine the solution for $\alpha=\begin{bmatrix}
        \alpha_0 & \alpha_1 & \cdots & \alpha_{q}
    \end{bmatrix}^\top$ that satisfies  
    \begin{equation}\label{eq:LGS-general-q}
        \underbrace{
            \begin{bmatrix}
                x_{q+1} \\
                x_{q+2} \\
                \vdots\\
                x_{q}
            \end{bmatrix}
        }_{\eqqcolon \tilde{b}}
        = 
        \underbrace{
            \begin{bmatrix}
                x_0 & x_1 & \cdots & x_{q} \\  
                x_1 & x_2 & \cdots & x_{q+1} \\
                \vdots & \vdots & \ddots & \vdots \\
                x_{p-2} & x_{0} & \cdots & x_{q-1}
            \end{bmatrix}
        }_{\eqqcolon \tilde{A}}
        \alpha,
    \end{equation}
    where we use the periodicity argument~\eqref{eq:periodicity-xk} for $k\in\bbN_{0:q}$.
    The Kronecker-Capelli theorem~\citep{kronecker:1903,capelli:1892} characterizes the existence of at least one solution if and only if the rank condition
    \begin{equation}\label{eq:Kronecker-Capelli}
        \rank(\tilde{A}) = \rank(\tilde{A}|\tilde{b}).
    \end{equation}
    holds. 
    Note that the sequence $\{x_k\}_{k=0}^{p-2}$ contains all values within $\bbN_{1:p-1}$ exactly once since $m$ functions as a generator of $\bbZ_p^*$.
    Moreover, we observe that the composite matrix $(\tilde{A}|\tilde{b})$ is composed of exactly this sequence organized in a Hankel matrix structure. 
    Hence,~\eqref{eq:Kronecker-Capelli} can only be satisfied for $q<p-2$ if there is a linear relation between elements in the sequence.
    \\
    To investigate this further, we leverage Corollary~\ref{cor:Euler-criterion-DH} to deduce
    \begin{equation}\label{eq:periodicity-mod-negative}
        x_{k+\frac{p-1}{2}} 
        = m^{\frac{p-1}{2}} x_k
        \overset{{\pmod{p}}}{=} 
        - x_k
        = p - x_k.
    \end{equation}
    Thus, we modify the system of equations in~\eqref{eq:LGS-general-q} and obtain%
    \begin{equation}\label{eq:LGS-general-q-Euler}%
        \begin{bmatrix}
            x_{q+1} \\
            x_{q+2} \\
            \vdots\\
            p-x_{q} \\
        \hline
            p-x_{q+1} \\
            p-x_{q+2} \\
            \vdots \\
            x_{q}
        \end{bmatrix}
        =
        \begin{bmatrix}
            x_0 & x_1 & \cdots & x_{q} \\  
            x_1 & x_2 & \cdots & x_{q+1} \\
            \vdots & \vdots & \ddots & \vdots \\
            x_{\frac{p-1}{2}-1} & p-x_0 & \cdots & p-x_{q-1} \\
        \hline
            p-x_0 & p-x_1 & \cdots & p-x_{q} \\
            p-x_{1} & p-x_{2} & \cdots & p-x_{q+1} \\
            \vdots & \vdots & \ddots & \vdots \\
            p-x_{\frac{p-1}{2}-1} & x_0 & \cdots & x_{q-1}
        \end{bmatrix}
        \alpha.
    \end{equation}%
    Importantly, the particular structure in~\eqref{eq:LGS-general-q-Euler} constraints $\alpha$ to satisfy $\sum_{j=0}^q \alpha_j = 1$. 
This can be easily observed by summing up, e.g., the first rows of the upper and lower block in~\eqref{eq:LGS-general-q-Euler}.
    Moreover, note that if the upper block of equations is satisfied for such an $\alpha$, the corresponding lower block is also satisfied.
    This allows us to simplify the system of equations by focusing only on the upper block and the additional constraint on $\alpha$, i.e., 
    \begin{equation}\label{eq:LGS-general-q-Euler-reduced}
        \begin{bmatrix}
            x_{q+1} \\
            x_{q+2} \\
            \vdots\\
            p-x_{q}\\
        \hline
            1
        \end{bmatrix}
        =
        \begin{bmatrix}
            x_0 & x_1 & \cdots & x_{q} \\  
            x_1 & x_2 & \cdots & x_{q+1} \\
            \vdots & \vdots & \ddots & \vdots \\
            x_{\frac{p-1}{2}-1} & p-x_0 & \cdots & p-x_{q-1} \\
        \hline
            1 & 1 & \cdots & 1
        \end{bmatrix}
        \alpha.
    \end{equation}
    Based on this condition we consider critical cases for the number $q$.
    \\
    \hspace*{5mm}\emph{Case $q=(p-1)/2-1=\tilde{q}-1$}.~
    Substituting the choice $q=\tilde{q}-1$ into~\eqref{eq:LGS-general-q-Euler-reduced} results in the simplified system of equations
    \begin{equation}\label{eq:LGS-general-q-Euler-reduced-full-rank}
        \underbrace{
            \begin{bmatrix}
                p-x_0 \\
                p-x_1 \\
                \vdots\\
                p-x_{\frac{p-1}{2}-1}\\
            \hline
                1
            \end{bmatrix}
        }_{\coloneqq \hat{b}}
        =
        \underbrace{
            \begin{bmatrix}
                x_0 & x_1 & \cdots & x_{\frac{p-1}{2}-1} \\  
                x_1 & x_2 & \cdots & p-x_0 \\
                \vdots & \vdots & \ddots & \vdots \\
                x_{\frac{p-1}{2}-1} & p-x_0 & \cdots & p-x_{\frac{p-1}{2}-2} \\
            \hline
                1 & 1 & \cdots & 1
            \end{bmatrix}
        }_{\coloneqq \hat{A}}
        \alpha.
    \end{equation}
    However, for this system of equations there exist no solution $\alpha$. 
    In particular, the sequence $\{x_k\}_{k=0}^{\frac{p-1}{2}}$ contains only distinct values since $m$ is a generator of $\bbZ_p^*$. 
    Moreover, $p-x_0=x_{\frac{p-1}{2}}$ according to~\eqref{eq:periodicity-mod-negative}.
    Due to the nonlinear modulo operation in the system dynamics~\eqref{eq:dynamical-system} and the last row containing only ones, we conclude that $\rank(\hat{A}|\hat{b}) = (p-1)/2+1$, i.e., $(\hat{A}|\hat{b})$ has full rank (see Appendix~\ref{sec:appendix:full-rank-Ab}). 
    Further, $\hat{A}$ can only have rank $(p-1)/2$ since this number is the number of columns of $\hat{A}$. 
    Hence,~\eqref{eq:Kronecker-Capelli} is violated, i.e., no solution exists according to the Kronecker-Capelli theorem.
    Thus, the selected value of $q$ is not large enough to obtain a linear system capable of capturing the nonlinear dynamics of the cryptosystem for all $k\in\bbN_0$.
    \\
    \hspace*{5mm}\emph{Case $q=(p-1)/2=\tilde{q}$}.~
    We proceed by substituting the choice $q=\tilde{q}$ into~\eqref{eq:LGS-general-q-Euler-reduced}.
    In particular, this yields the system of equations 
    \begin{equation}\label{eq:LGS-periodicity-mod-negative-p}
        \begin{bmatrix}
            p-x_{1} \\
            p-x_{2} \\
            \vdots\\
            x_{0} \\
        \hline
            1
        \end{bmatrix}
        = \begin{bmatrix}
            x_0 & x_1 & \cdots & p-x_{0} \\  
            x_1 & x_2 & \cdots & p-x_{1} \\
            \vdots & \vdots & \ddots & \vdots \\
            x_{\frac{p-1}{2}-1} & p-x_0 & \cdots & p-x_{\frac{p-1}{2}-1} \\
        \hline
            1 & 1 & \cdots & 1
        \end{bmatrix}
        \alpha.
    \end{equation}
    Note that~\eqref{eq:LGS-periodicity-mod-negative-p} consists of the matrix $(\hat{A}|\hat{b})$ on the right-hand side. 
    Due to its full rank, we conclude that the system of equations has a solution. 
    In particular,~\eqref{eq:LGS-periodicity-mod-negative-p} is satisfied for $\alpha=\begin{bmatrix} 1 & -1 & 0 & \cdots & 0 & 1 \end{bmatrix}^\top$. 
    Hence, the lifted linear system $z_{k+1} = A z_k$ with companion matrix $A$ corresponding to the derived $\alpha$ describes the nonlinear cryptosystem for all $k\in\bbN_0$. 
    This conclusion is valid for $\tilde{q}=(p-1)/2$ and the inclusion of $\tilde{q}+1$ observables, such that the statement of the theorem is established.
\end{pf}
\vspace*{-\baselineskip}
In order to illustrate Theorem~\ref{thm:minimal-number-observables-DH} graphically, we consider the following example.
\begin{exmp}\label{exmp:DH-example}
    Let the public modulus $p$ and a corresponding generator $m\in\bbZ_p^*$ of a DH cryptosystem be given as $p=19$ and $m=2$. 
    The trajectory of the corresponding nonlinear system~\eqref{eq:dynamical-system} is depicted in Fig.~\ref{fig:example-DH}, where its periodicity is clearly visible since $x_{18}=1=x_0$.
    To find a linear representation of the DH cryptosystem, we rely on the Koopman representation described in Section~\ref{sec:observables-value-list}.
    To this end, we need to specify the necessary number of lifting functions. 
    In particular, Theorem~\ref{thm:minimal-number-observables-DH} yields the minimal lifting dimension $\tilde{q}+1$ with $\tilde{q} = (p-1)/2=9$.
    This number can also be graphically confirmed since the trajectory can be divided into two parts. 
    First, the behavior of the trajectory is random-like without a recognizable structure. 
    However, the second part starting at $k=9$ follows indeed a structure and can be related to the first part. 
    More specifically, the second part of the trajectory is just the first part horizontally mirrored at zero and shifted by $p=19$.
    This is in line with the relation discovered in~\eqref{eq:periodicity-mod-negative}, where $x_k$ can be related to $x_{k+\tilde{q}}=p-x_k$.
    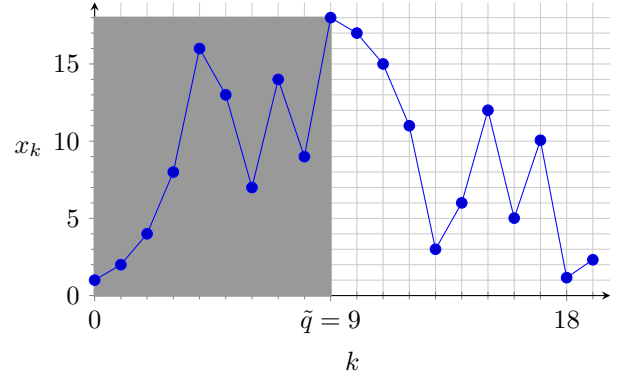
\begin{figure}[tb]
        \centering
        \begin{tikzpicture}[%
            /pgfplots/every axis y label/.style={at={(0,0.5)},xshift=-25pt},%
            ]%
            \begin{axis}[%
                width=\columnwidth,
                height=0.65\columnwidth,
                axis x line=bottom,
                x axis line style=-{stealth},
                axis y line=left,
                xmin=0,xmax=19.66,
                ymin=0,ymax=18.99,
                xtick={0,9,18},
                xticklabels={0,$\tilde{q}=9$,18},
                ylabel={$x_k$},
                xlabel={$k$},
                samples at={0,...,19},
                clip=false,
                grid=both,
                major grid style={gray!40},
                minor xtick={1,2,3,4,5,6,7,8,10,11,12,13,14,15,16,17,19},
                minor y tick num=4,
                minor grid style={gray!40},
            ]
                \filldraw[thick,gray!80,opacity=0.5] (axis cs:0,0) -- (axis cs:9,0) -- (axis cs:9,18) -- (axis cs:0,18) -- cycle;
                \addplot[black!75,mark=*] {mod(2^x, 19)};
                \label{fig:plot:trajectory}
            \end{axis}
        \end{tikzpicture}
        \vspace*{-0.6\baselineskip}
        \caption{Trajectory of~\eqref{eq:dynamical-system} for $p=19$ and $m=2$ corresponding to a DH cryptosystem~\eqref{fig:plot:trajectory}. The gray area highlights the minimal number of values in the state of the linear system representation according to Theorem~\ref{thm:minimal-number-observables-DH}.}
        \label{fig:example-DH}
    \end{figure}%
\end{exmp}%
\vspace*{-0.5\baselineskip}
\begin{rem}
    The observables on the complex unit circle show a similar behavior as for a value list seen in Example~\ref{exmp:DH-example} and Fig.~\ref{fig:example-DH}. In particular, Fig.~\ref{fig:observables-complex-unit-circle} exemplarily visualizing the observables in Section~\ref{sec:observables-complex} indicates that the observables are symmetric w.r.t. the real axis.
\end{rem}
\begin{cor}[{\citeauthor{schlor:strasser:allgower:2023},~\citeyear{schlor:strasser:allgower:2023},~Cor.~8}]\label{cor:linear-system-for-q}%
    Let $z=h(x)$, $h:\bbN\to\bbC^{q+1}$, with $q\geq \tilde{q}$ define a lifted state of the nonlinear DH cryptosystem according to Section~\ref{sec:observables-value-list}. Then, the system $z_{k+1} = \bar{A} z_k$ is an equivalent representation of~\eqref{eq:dynamical-system}, where $\bar{A}$ is a companion matrix with $\alpha_{q-\tilde{q}} = 1$, $\alpha_{q-\tilde{q}+1} = -1$, $\alpha_{q} = 1$ and $\alpha_j = 0$ for $j\in\bbN_{0:q-\tilde{q}-1}\cup\bbN_{q-\tilde{q}+2:q-1}$.
\end{cor}%
Corollary~\ref{cor:linear-system-for-q} aligns also with the representation in~\eqref{eq:dynamics-linear-system-upper-bound} for $q=p-2 = 2\tilde{q}-1$. This correspondence gets clear when we analyze
\begin{align*}
    x_{p-1} &= x_{\tilde{q}-1} - x_{\tilde{q}} + x_{p-2} \\
    &= x_{\tilde{q}-1} - (p-x_0) + (p-x_{\tilde{q}-1}) 
    = x_0,
\end{align*}
where we use~\eqref{eq:periodicity-mod-negative} for the second equation.

Given that the state matrix $A$ in the resulting linear representation $z_{k+1} = A z_k$ takes the form of a companion matrix, it follows that all eigenvalues correspond to roots of the associated polynomial $p(\mu) = \mu^{q+1} - \sum_{j=0}^{q} \alpha_j\mu^j$.
To be more precise, this can be stated as
\begin{equation*}
    0 = p(\mu) = \mu^{q+1} - \mu^q + \mu - 1 = (\mu^{q} + 1)(\mu - 1).
\end{equation*}
As a result, the companion matrix $A$ always has the eigenvalue $\mu_1 = 1$. 
Further, all eigenvalues are part of the complex unit circle and distinct since $m<p$.
This fact ensures diagonalizability of $A$ and justifies disregarding the eigenvalues' absolute values in the reconstruction of $e$, as discussed in Section~\ref{sec:reconstructE}.
\\
The obtained $\tilde{q}=(p-1)/2$ corresponds to an observable vector of dimension $\tilde{q}+1$. 
If $p-1$ is not divisible by $4$, which implies that $\tilde{q}$ is odd, we can deduce that the lifted state dimension $\tilde{q}+1$ is even. 
For any odd $q$, it is easy to see that $\mu_{\tilde{k}}=-1$ is an eigenvalue of $A$ since $\mu_{\tilde{k}}^{q} + 1 = 0$.
Hence, we can determine if the secret number $e$ is an even or odd number by evaluating 
\begin{equation*}
    \frac{\tilde{z}_{e,\tilde{k}}}{\tilde{z}_{0,\tilde{k}}} =
    \begin{cases}
        1 & \text{if $e$ is even,} \\
        -1 & \text{if $e$ is odd,}
    \end{cases}
\end{equation*}
where $\tilde{z}_{e,\tilde{k}}$ and $\tilde{z}_{0,\tilde{k}}$ are defined in~\eqref{eq:z_e-tilde}, and $\tilde{k}\in\bbN_{0:q}$ corresponds to the eigenvalue $\mu_{\tilde{k}}=-1$.
It is important to emphasize that this analysis holds true only if $-1$ is indeed an eigenvalue of $A$, specifically when $q$ is an odd number.
In contrast, Section~\ref{sec:reconstructE} presents a more comprehensive exploration of reconstructing $e$ without confining $q$ to being odd.
\\
Further, if the complete list of values for $x_k$ is contained within the observable vector, i.e., $q=p-2$, an alternative method for reconstructing $e$ through the structure of $z=h(x)$ is possible.
This approach does not follow the discussion in Section~\ref{sec:reconstructE} but exploits the fact that $z_0=h(x_0)$ includes all $p-1$ attainable system values arranged in sequential order. 
In particular, $x_0$ is at the first position and $x_e$ is at the $(e+1)$-th position within $z_0$. 
Hence, the brute-force reconstruction investigates which element of $z_0$ corresponds to the measured value $x_e$.
\\
The overall procedure reconstructing the secret integer $e$ is summarized in Algorithm~\ref{alg:DH}.
\begin{algorithm}[tb]
    \caption{Reconstruction of secret integer $e$ for DH}\label{alg:DH}
    \textbf{Input:} prime number $p$, $m\in\bbZ_p^*$, public ciphertext $c_e$\\
    \textbf{Output:} secret integer $e$\\[0.5ex]
    \emph{Initialize the algorithm:}
    \begin{spacing}{0.85}
	    \begin{algorithmic}
	        \State Define the companion matrix in~\eqref{eq:Koopman-A-matrix-RNF} for $q=\frac{p-1}{2}$ and $\alpha=\begin{bmatrix} 
	                1 & -1 & 0 & \cdots & 0 & 1
	            \end{bmatrix}^\top$
	        \State Calculate the eigendecomposition $A=V\Lambda V^{-1}$
	        \State Define $\tilde{z}_0$ and $\tilde{z}_e$ according to~\eqref{eq:z_e-tilde}
	    \end{algorithmic}
    \end{spacing}
    \vspace*{-\baselineskip}~\\[0.5ex]
    \emph{Determine if $e$ is even or odd:}
    \begin{spacing}{0.85}
	    \begin{algorithmic}
	        \If{$q+1$ is even}
	            \State Find $\tilde{k}$ corresponding to $\mu_{\tilde{k}}=-1$
	            \If{$\operatorname{sgn}(\tilde{z}_{0,\tilde{k}})$ equals $\operatorname{sgn}(\tilde{z}_{e,\tilde{k}})$}
	                \State $e$ is even
	            \Else
	                \State $e$ is odd
	            \EndIf
	        \EndIf
	    \end{algorithmic}
	\end{spacing}
    \vspace*{-\baselineskip}~\\[0.5ex]
    \emph{Reconstruction of $e$ based on complex angles:}
    \begin{spacing}{0.85}
	    \begin{algorithmic}
	        \Statex Define $\kappa=\{k\in\bbN_{0:q}\mid\mu_k\in\bbC\setminus\bbR\}$, $\tilde{\Lambda} = \diag(\Lambda)$, and $[\hat{\Lambda},\hat{z}_0,\hat{z}_e] = [\tilde{\Lambda}_\kappa,\tilde{z}_{0,\kappa},\tilde{z}_{e,\kappa}]$
	        \Statex $a \gets \operatorname{angle}(\hat{\Lambda})$
	        \Statex $b \gets \operatorname{angle}(\hat{V}_e \oslash \hat{V}_0)$
	        \If{$b$ is zero}
	            \State $c \gets 0$
	        \Else 
	            \State $a \gets a \pmod{2\pi}$
	            \State $b \gets b \pmod{2\pi}$
	            \State $c \gets b \oslash a$
	            \While{$c$ contains not only integers}
	                \State $b \gets b + 2\pi$
	                \State $c \gets b \oslash a$
	            \EndWhile
	            \State $e \gets \max(c \mod{\lambda(p)})$
	        \EndIf
	    \end{algorithmic}
 	\end{spacing}
\end{algorithm}
Table~\ref{tab:DH-computation-times} shows the computation time of the algorithm for different prime numbers $p$, where we list both the worst-case and the average computation time over all combinations of $p$ with possible generators $m\in\bbZ_p^*$.
\bgroup
\renewcommand{\figurename}{Tab.} 
\renewcommand{\thefigure}{1}
\renewcommand{\arraystretch}{0.75}%
\begin{figure}[tb]
    \centering
    \caption{Computation times of Algorithm~\ref{alg:DH}.}
    \label{tab:DH-computation-times}
    \renewcommand{\tabcolsep}{5pt}
    \begin{tabular}{r|r|r}
        \toprule
        $p$\phantom{97}& worst-case time [s] & average time [s] \\
        \midrule
        97 & 0.0575 & 0.0096 \\
        997 & 1.0611 & 0.4274 \\
        9973 & 126.8642 & 101.6167 \\
        31607 & 3937.8940 & 3561.7069 \\
        \bottomrule
    \end{tabular}
\end{figure}
\egroup
The computation times are obtained by running the algorithm on a standard laptop with an Intel Core i7-10510U CPU and 32 GB of RAM. 
Larger prime numbers are computationally more demanding, which is reflected in the computation times, and the required memory did only allow us to run the algorithm for prime numbers $p\in\cO(10^{4.5})$.
We note that while there exist more efficient algorithms for attacking DH in the literature, see, e.g.,~\cite{shoup:1997}, our approach allows to bridge the fields of Koopman theory and cryptography.

\vspace*{-0.35\baselineskip}
\subsection{Minimal number of observables for RSA}\label{sec:minimal-observable-number-RSA}
\vspace*{-0.35\baselineskip}
In this section, we establish an upper bound on the minimal number of observables for a linear representation of RSA. The derived results build on the insights gained in Section~\ref{sec:minimal-observable-number-DH} for DH. However, the derivation requires additional number theoretical arguments since the modulus $p$ is no prime number but a product of two prime numbers. Thus, we generalize Euler's criterion first.
\begin{lem}[Generalized Euler's criterion]\label{lm:Euler-criterion-generalization}
    If $p=p_1p_2$ is a product of an odd prime number $p_1$ and $p_2\neq p_1$, which is either an odd prime number or $p_2=1$, then for any integer $m$ coprime to $p$ holds
    \begin{equation}\label{eq:Euler-criterion-generalization}
        m^{\frac{\varPhi(p)}{2}} \overset{{\pmod{p}}}{=} \begin{cases}
            1 & \text{if}\>m\,\mathrm{R}\,p \>\lor\> p_2\neq 1, \\
            -1 & \text{else}.
        \end{cases}
    \end{equation}
\end{lem}
\vspace*{-\baselineskip}
\begin{pf}
    See Appendix~\ref{sec:appendix:Euler-criterion-generalization-proof}.
\end{pf}
\vspace*{-\baselineskip}
In~\cite{ohnari:1977}, the statement of Lemma~\ref{lm:Euler-criterion-generalization} was shown for $p_2 = 1$ and $p$ chosen as $2^2$, $p_1^r$, or $2p_1^r$ for an odd prime number $p_1$ and a positive integer $r$. To the best of our knowledge, this work is the first applying Euler's criterion to a modulus $p$ which is the product of two distinct odd prime numbers. While this result is of its own interest, we modify it in the following to determine the minimal required lifting dimension for a linear representation of RSA. To this end, we recall
\begin{equation}\label{eq:Carmichael}
    m^{\lambda(p)} \pmod{p} = 1
\end{equation}
for any $m$ coprime to $p$ due to the definition of the Carmichael function.
\begin{cor}\label{cor:Carmichael-criterion}
    Any $p=p_1p_2$ with two distinct odd prime numbers $p_1,p_2$ and any integer $m\in\bbZ_p^*$ corresponding to RSA satisfy
    \begin{equation*}
        m^{\lambda(p)} \pmod{p}
        = 1
        = m^{\frac{\varPhi(p)}{2}} \pmod{p}
    \end{equation*}
    with $\lambda(p)\leq \frac{\varPhi(p)}{2}$.
\end{cor}
\vspace*{-\baselineskip}
\begin{pf}
    The first identity results trivially due to the definition of the Carmichael function in~\eqref{eq:Carmichael} since $m$ and $p$ are coprime. 
    Let $p=p_1p_2$ be the product of two distinct odd prime numbers $p_1$ and $p_2$, i.e., $p_2\neq 1$. Thus, using Lemma~\ref{lm:Euler-criterion-generalization} directly yields the second identity.
    \\
    It remains to show that $\lambda(p)\leq \frac{\varPhi(p)}{2}$.
    Clearly, $\lambda(p)=\mathrm{lcm}(\lambda(p_1),\lambda(p_2)) \leq \lambda(p_1)\lambda(p_2)$. As both $p_1$ and $p_2$ are odd prime numbers, we know that both $\lambda(p_1) = \varPhi(p_1) = p_1 - 1$ and $\lambda(p_2) = \varPhi(p_2) = p_2 - 1$ are even. Hence, the least common multiple can be reduced by at least the factor $2$ compared to the product $\varPhi(p)=\lambda(p_1)\lambda(p_2)$ and we deduce
    $
        \lambda(p)
        \leq \lambda(p_1)\lambda(p_2)/2
        = \varPhi(p)/2
    $.
\end{pf}
\vspace*{-\baselineskip}
Now we can state our second main theorem which is related to Theorem~\ref{thm:minimal-number-observables-DH}, but $p$ is not restricted to prime numbers enabling the analysis of RSA cryptosystems.
\begin{thm}\label{thm:minimal-number-observables-RSA}
    Let $p=p_1p_2$ be the product of two distinct odd prime numbers $p_1$ and $p_2$. Then, an upper bound for the minimal lifting dimension to obtain a linear representation of the RSA cryptosystem for any $m\in\bbZ_p^*$ with the observables in Section~\ref{sec:observables-value-list} is $\tilde{q}+1$, where $\tilde{q}=\lambda(p)-1$.
\end{thm}
\vspace*{-\baselineskip}
\begin{pf}
    We start the proof by investigating the linear representation~\eqref{eq:linear-dynamics-general-q} of a cryptosystem described by~\eqref{eq:dynamical-system} for a general $q\in\bbN$. As before, the parameter vector $\alpha$ needs to be chosen such that~\eqref{eq:closing-condition} holds for all $k\in\bbN_0$. Recall that $x_{k+\varPhi(p)}=x_k$ due to Lemma~\ref{lm:Euler}. This relation can be tightened using Corollary~\ref{cor:Carmichael-criterion} leading to 
    \begin{alignat}{2}
        x_k
        &= m^{\frac{\varPhi(p)}{2}} x_k \pmod{p}& 
        &= x_{k+\frac{\varPhi(p)}{2}} x_k 
        \nonumber
        \\
        &= m^{\lambda(p)} x_k \pmod{p} &
        &= x_{k+\lambda(p)},
        \label{eq:periodicity-xk-RSA}
    \end{alignat}
    where $x_{k+\lambda(p)}$ corresponds to a (possibly) earlier time index than $x_{k+\frac{\varPhi(p)}{2}}$ due to Corollary~\ref{cor:Carmichael-criterion}.
    Thus, we only need to consider $k\in\bbN_{0:\lambda(p)-1}$. In particular, we solve for $\alpha=\begin{bmatrix}
        \alpha_0 & \alpha_1 & \cdots & \alpha_{q}
    \end{bmatrix}^\top$ satisfying
    \begin{equation}\label{eq:LGS-general-q-RSA}
        \underbrace{
            \begin{bmatrix}
                x_{q+1} \\
                x_{q+2} \\
                \vdots\\
                x_{q}
            \end{bmatrix}
        }_{\eqqcolon \tilde{b}}
        = 
        \underbrace{
            \begin{bmatrix}
                x_0 & x_1 & \cdots & x_{q} \\  
                x_1 & x_2 & \cdots & x_{q+1} \\
                \vdots & \vdots & \ddots & \vdots \\
                x_{\lambda(p)-1} & x_{0} & \cdots & x_{q-1}
            \end{bmatrix}
        }_{\eqqcolon \tilde{A}}
        \alpha,
    \end{equation}
    where we use~\eqref{eq:periodicity-xk-RSA} for $k\in\bbN_{0:q}$. 
    \\
    Note that $(\tilde{A}|\tilde{b})$ consist of the sequence $\{x_k\}_{k=0}^{\lambda(p)-1}$ arranged in a Hankel matrix.
    Moreover, for every $p$ there exists a value $m$ such that the values $x_k$, $k\in\bbN_{0:\lambda(p)-1}$, are all different due to the definition of $\lambda(p)$. 
    Thus, the rank condition~\eqref{eq:Kronecker-Capelli} yields that the minimal required $q$ is less or equal to $\lambda(p)-1$. 
    Further, $q<\lambda(p)-1$ is only possible if some columns in $(\tilde{A}|\tilde{b})$ are linearly dependent.
    \\
    Using the upper bound $q= \lambda(p)-1=\tilde{q}$ in~\eqref{eq:LGS-general-q-RSA} together with~\eqref{eq:periodicity-xk-RSA} results in the system of equations
    \begin{equation*}
        \begin{bmatrix}
            x_0 \\
            x_1 \\
            \vdots\\
            x_{\lambda(p)-1}
        \end{bmatrix}
        = \begin{bmatrix}
            x_0 & x_1 & \cdots & x_{\lambda(p)-1} \\  
            x_1 & x_2 & \cdots & x_0 \\
            \vdots & \vdots & \ddots & \vdots \\
            x_{\lambda(p)-1} & x_0 & \cdots & x_{\lambda(p)-2}
        \end{bmatrix}
        \alpha,
    \end{equation*}
    which is satisfied for $\alpha=\begin{bmatrix} 1 & 0 & \cdots & 0\end{bmatrix}^\top$.         
    Hence, the nonlinear cryptosystem is fully characterized by the lifted linear system $z_{k+1} = A z_k$ for the matrix $A$ with the derived $\alpha$ for $\tilde{q} = \lambda(p) - 1$ and $\tilde{q}+1 = \lambda(p)$ observables.
\end{pf}
\vspace*{-\baselineskip}
Similar to the discussion in Section~\ref{sec:minimal-observable-number-DH}, if the lifting dimension $q$ is odd, then we can analyze the eigenvalues of the state matrix $A$ to deduce whether the secret key $d$ is even or odd. Otherwise, we can still apply the more detailed investigation of $d$ in Section~\ref{sec:reconstructD} which reconstructs $d$ without restriction to odd numbers $q$. 
For RSA, the matrix $A$ is a shift matrix, i.e., $A\in\bbR^{q+1\times q+1}$ has the distinct eigenvalues $\mu_j = \exp\left(i \frac{2\pi}{q+1}j\right)$ for $j=0,1,...,q$. Thus, $A$ is diagonalizable and all eigenvalues lie on the complex unit circle, where $\mu_0 = 1$ is always an eigenvalue of $A$.
\\
For the derived $q=\lambda(p)-1$ we get an observable vector of dimension $\lambda(p)$. Since $\lambda(p)=\operatorname{lcm}(p_1-1,p_2-1)$ is always even for odd prime numbers $p_1,p_2$, we deduce that also the lifted state dimension is even. Hence, it is easy to see that $\mu_{\frac{\lambda(p)}{2}} = -1$ is an eigenvalue of $A$.
As a consequence, we can state whether the secret key $d$ is even or odd via
\begin{equation*}
    \frac{\tilde{z}_{d,\frac{\lambda(p)}{2}}}{\tilde{z}_{0,\frac{\lambda(p)}{2}}} =
    \begin{cases}
        1 & \text{if $d$ is even,} \\
        -1 & \text{if $d$ is odd,}
    \end{cases}
\end{equation*}
with $\tilde{z}_{d,\frac{\lambda(p)}{2}},\tilde{z}_{0,k}$ defined in~\eqref{eq:z_e-tilde} for $k=\frac{\lambda(p)}{2}$.
\\
We note that the reconstruction of the secret key $d$ based on the derived linear representation of the RSA cryptosystem follows again the procedure in Algorithm~\ref{alg:DH}. This is in line with the discussion in Section~\ref{sec:reconstructD}. The only modification in order to obtain the secret key $d$ as the output of the algorithm is the usage of a product of two distinct prime numbers $p=p_1p_2$, a message $m$, and its corresponding ciphertext $c$ as inputs with the derived $q=\lambda(p)-1$ and corresponding vector $\alpha$.
\begin{rem}
    The result on the minimal lifting dimension for the case of RSA in Theorem~\ref{thm:minimal-number-observables-RSA} holds true for \emph{any} $m\in\bbZ_p^*$. 
    However, for a given message $m$ and modulus $p$ there might exist a smaller period length $\zeta(m)\leq\lambda(p)$ for which $x_{k+\zeta} = x_k$.
    Thus, it would suffice to use only a lifting of dimension $\zeta(m)$ for this specific combination of $m$ and $p$ instead of using a lifting of dimension $\tilde{q}+1$, but the lifting would depend explicitly on the message $m$.
    Moreover, even though this possibly smaller period length $\zeta(m)$ is typically unavailable, the proposed Algorithm~\ref{alg:DH} recovers the true (unknown) secret key $d$ as $\hat{d} = d \pmod{\zeta(m)}$.
    \\
    Moreover, there always exists a message $m$ for a given $p$ such that the bound in Theorem~\ref{thm:minimal-number-observables-RSA} is tight. 
    For example, given $p=p_1p_2$ with $p_1=3$ and $p_2=5$, the derived upper bound yields $\tilde{q}=3$. This is indeed the smallest possible lifting dimension such that there exists a linear system representation based on the observables in Section~\ref{sec:observables-value-list} for, e.g., $m=2$.
    Contrary, $m=4$ leads to linear dependencies in a lifting of dimension $\tilde{q}+1$ and, thus, a lifting of dimension $\check{q}+1$ with $\check{q}=1$ is already sufficient to obtain a linear representation.
\end{rem}

\vspace*{-0.35\baselineskip}
\subsection{Relation to Willems' Fundamental Lemma}
\vspace*{-0.35\baselineskip}
According to~\eqref{eq:linear-dynamics-general-q}, the resulting lifted dynamics at time $k=0$ for a chosen $q$ corresponding to data $\{x_k\}_{k=0}^{2q+1}$ is exact for the next $q$ steps if $\alpha$ satisfies  
\begin{equation*}
    \begin{bmatrix}
        x_{q+1} \\
        x_{q+2} \\
        \vdots \\
        x_{2q+1}
    \end{bmatrix}
    = \begin{bmatrix}
        x_{0} & x_{1} & \cdots & x_{q} \\
        x_{1} & x_{2} & \cdots & x_{q+1} \\
        \vdots & \vdots & \ddots & \vdots \\
        x_{q} & x_{q+1} & \cdots & x_{2q}
    \end{bmatrix}
    \alpha.
\end{equation*}
Note that the system of equations is described by a Hankel matrix of order $q+1$, which we denote by $H_{q+1}(x)$. A solution $\alpha$ exists if the rank condition in~\eqref{eq:Kronecker-Capelli} holds. 

\begin{rem}
    Suppose $\{x_k\}_{k=0}^{N-1}$ is a trajectory of a system which can be described by a linear time-invariant (LTI) system. Then, $\{\bar{x}_k\}_{k=0}^{L-1}$ is a trajectory of this LTI system if and only if there exists $\alpha\in\bbR^{N-L+1}$ such that 
    $
        H_{L}(x) \alpha = \bar{x}
    $.
    This is a direct application of Willems' Fundamental Lemma~\citep{willems:rapisarda:markovsky:demoor:2005} to discrete-valued autonomous systems. To put it in other words, we only find an LTI system generating a trajectory $\bar{x}$ of length $L$ iff there exists an $\alpha\in\bbR^{N-L+1}$ such that $H_{L}(x) \alpha = \bar{x}$ for measured data $x$ of length $N$. For $N=2q$ and $L=q+1$ we obtain an $\alpha$ satisfying~\eqref{eq:linear-dynamics-general-q} for $k\in\bbN_{0:q}$. 
    \\
    Since our underlying system for DH and RSA is periodic with period $\lambda(p)$, we want to generate trajectories of length $\lambda(p)$ which are repeated afterwards due to the periodicity. Thus, we require $L=\lambda(p)$ and hence, $q=\lambda(p)-1,N=2\lambda(p)-1$. Since the values for $k\geq\lambda(p)$ are just periodically repeated values of the first $\lambda(p)$ values, it is sufficient to have a trajectory of length $N=\lambda(p)$.
    \\
    For DH, we choose, according to Theorem~\ref{thm:minimal-number-observables-DH}, $q=\lambda(p)/2$ and reduce the trajectory length to $N=\lambda(p)/2$ while still generating trajectories of length $\lambda(p)$ and above.
\end{rem}

\vspace*{-0.35\baselineskip}
\subsection{Relation to linear complexity}\label{sec:linear_complexity}
\vspace*{-0.35\baselineskip}
The minimal number of states within the linear representation is closely linked to the linear complexity of the sequence it generates, a concept introduced in~\cite{beth:dai:1990, wang:1999}.%
\begin{defn}[Linear complexity]%
    The linear complexity of a sequence is defined as the length of the shortest linear feedback shift register (LFSR) required to generate the sequence.
\end{defn}%
An LFSR constitutes an array of values where each element shifts during every time step, and the subsequent value is generated by forming a linear combination of the existing entries.
Essentially, an LFSR embodies a linear dynamic system realized in companion form, which is the identical structure as emerges in Theorem~\ref{thm:minimal-number-observables-DH} and Theorem~\ref{thm:minimal-number-observables-RSA} for the observables in Section~\ref{sec:observables-value-list}. 
Thus, the linear complexity of the system's trajectory establishes an upper bound on the smallest state dimension for the higher-dimensional linear system representation.
This upper bound is reached in some cases when the observables are selected as a linear function of the generated sequence, related to what is discussed in Section~\ref{sec:minimal-observable-number-DH} for the DH cryptosystem.
\\
The Berlekamp-Massey algorithm (BMA)~\citep{berlekamp:1966,massey:1969} provides a way to determine the LFSR with the shortest length. 
However, our method of deriving a linear system using Koopman theory is more versatile than the LFSR approach and could potentially yield a linear system with a smaller state dimension.
While a comparison to general finite-state machines is left open for future work, this fact becomes easily obvious in at least two aspects.
First, the observable function $h$ offers the flexibility to incorporate nonlinearity by establishing a nonlinear dependence on the system's state. 
In particular, this entails the potential for a nonlinear dependence not only on the present state but also on past and predicted states. 
In contrast, the LFSR restricts itself to a linear combination of these states.
The potential for a reduction in state dimensionality is demonstrated in the subsequent example.
\begin{exmp}[{\citeauthor{schlor:strasser:allgower:2023},~\citeyear{schlor:strasser:allgower:2023},~Exmp.~11}]\label{exmp:scalar}
    Consider the sequence $\{0,1,2,0,1,2,...\}$ resulting from the dynamical system
    $
        x_{k+1} = x_k + 1 \pmod{3}
    $
    with
    $ 
        x_0 = 0
    $.
    This system is nonlinear due to the modulo operation.
    However, we can transform it into a linear representation by introducing the lifted state $y_k = (x_k,x_{k+1},x_{k+2})$ of higher dimension. 
    In particular, the lifted dynamics read 
    \begin{equation*}
        y_{k+1} 
        = \begin{bmatrix}
            x_{k+1} \\ x_{k+2} \\ x_{k+3}
        \end{bmatrix}
        = \begin{bmatrix}
            x_{k+1} \\ x_{k+2} \\ x_{k}
        \end{bmatrix}
        = \begin{bmatrix}
            0 & 1 & 0 \\
            0 & 0 & 1 \\
            1 & 0 & 0
        \end{bmatrix}
        y_k.
    \end{equation*}
    However, this straightforward lifting results in a lifting dimension that is as large as the number of distinct elements in the sequence. 
    Consequently, the linear complexity of the generated sequence is three, which is computed via the BMA and the LFSR $x_{k}=x_{k-3}$ with $x_0=0$, $x_1=1$, $x_2=2$.
    Alternatively, employing the Koopman approach, we leverage the periodicity of the system's dynamics by mapping them to the complex unit circle. This is achieved by the lifted state $z_k=\exp\left({i \tfrac{2\pi}{3} x_k}\right)$ with   
    \begin{equation*}
        z_{k+1} = \exp\left({i \tfrac{2\pi}{3} x_{k+1}}\right) 
        = \exp\left({i \tfrac{2\pi}{3} \left(x_k+1\pmod{3}\right)}\right).
    \end{equation*}
    Using the fact that $\xi \pmod{3} = \xi + 3 \ell$ for some integer $\ell\in\bbZ$, we derive a linear representation that generates the sequence via
    \begin{align*}
        z_{k+1} &= \exp\left({i \tfrac{2\pi}{3} (x_k + 1) + i \tfrac{2\pi}{3}3\ell}\right) 
        \\
        &= \exp\left({i\tfrac{2\pi}{3}}\right) \exp\left({i\tfrac{2\pi}{3}}\right)x_k
        = \exp\left({i\tfrac{2\pi}{3}}\right) z_k.
    \end{align*}
    Further, we express $x_k$ in terms of the lifted state $z_k$ as
    $
        x_k = - i \tfrac{3}{2\pi} \ln(z_k) \pmod{3}
    $.
    This Koopman lifting approach offers a linear system with a smaller state dimension than the corresponding LFSR.
\end{exmp}%
Second, within our linear representation, the state has the capacity to contain additional beneficial information, e.g., parameters of the nonlinear map. 
In contrast, the state of an LFSR is restricted to elements of the sequence exclusively. 
The subsequent examples illustrate the advantages of integrating this type of data.
\begin{exmp}\label{exmp:affine_normalized}
    Consider the sequence $\{0,2,4,6,8,10,...\}$, which can be described by 
    $
        x_{k+1} = x_k + 2
    $
    with 
    $  
        x_0=0
    $.
    Note that this system is not linear but affine. We choose a nonlinear transformation $h(x) = \exp({x})$ and define $z_k = h(x_k)$. Then, we can generate the sequence via 
    \begin{align*}
        z_{k+1} &= \exp({x_{k+1}}) 
        = \exp({x_k + 2}), 
        \qquad 
        z_0 = \exp(0) = 1,\\
        &= \exp(2) \exp({x_k}) 
        = \exp(2) z_k,
    \end{align*}
    where the dynamics of $z$ are linear and $x_{k} = \ln(z_k)$. 
    Using the BMA instead, we obtain the LFSR $x_{k}=-x_{k-2} + 2 x_{k-1}$ with $x_0=0$, $x_1=2$ and linear complexity two.
\end{exmp}
\begin{exmp}[{\citeauthor{schlor:strasser:allgower:2023},~\citeyear{schlor:strasser:allgower:2023},~Exmp.~12}]\label{exmp:affine}
    Consider the sequence $\{1,4,10,22,46,...\}$, which is governed by the system dynamics
    $
        x_{k+1} = m x_k + a
    $
    with 
    $
        x_0 = 1
    $,
    where $m=2$, $a=2$.
    The technique employed in Example~\ref{exmp:affine_normalized} cannot be applied here since $m\neq 1$. 
    Instead, we can express the system alternatively via linear dynamics of dimension two with the lifted state $z_k=(x_k,a)$.
    This transformation leads to the dynamics 
    \begin{equation*}
        z_{k+1} = \begin{bmatrix}
            m & 1 \\ 0 & 1
        \end{bmatrix}
        z_k, 
        \quad 
        z_0 = \begin{bmatrix}
            x_0 \\ a
        \end{bmatrix},
        \quad
        x_k = \begin{bmatrix}
            1 & 0
        \end{bmatrix} 
        z_k.
    \end{equation*}
    Note that this representation requires a higher state dimension compared to the original nonlinear system.
    However, the advantage is that the established representation is \emph{linear}.
    Despite the increased dimensionality, the state dimension remains significantly smaller than that of the LFSR $x_{k}=3x_{k-1}-2x_{k-2}$ with $x_0=1$, $x_1=4$ needed to generate the sequence, which has linear complexity two.
\end{exmp}
\begin{exmp}
    Consider the sequence $\{2,4,16,256,...\}$, which can be described by 
    $
        x_{k+1} = m x_k^b
    $
    with 
    $
        x_0 = 2
    $,
    where $m\in\bbN$ and $b\in\bbN$. In particular, the sequence is generated with $m=1$ and $b=2$. Here, the BMA cannot find a corresponding LFSR. 
    Instead, we choose the nonlinear transformation $h(x) = \ln(x)$ and define $z_k = h(x_k)$. The transformation is well-defined since $m\in\bbN$ and, thus, $x_k>0$ for all $k\geq 0$. Then, we explain the sequence via 
    \begin{equation*}
        z_{k+1} = \ln(x_{k+1}) = \ln(m x_k^b) = b z_k + \ln(m)
    \end{equation*}
    with $z_0 = \ln(x_0)$, where the dynamics of $z$ are affine and $x_{k} = \exp({z_k})$. 
    Note that the dynamics are linear for $m=1$. For $m > 1$ we can proceed as in Example~\ref{exmp:affine}. 
\end{exmp}
Since the linear complexity of any sequence is smaller or equal to the length of the sequence, we always find a linear representation which is smaller or equal to the number of elements in the sequence.
This is in agreement with the results in Section~\ref{sec:minimal-observable-number}.

\begin{rem} 
    Note that the number of elementary operations with an integer $x\in\bbZ$ is limited.
    The integers $\bbZ$ are closed under the operations of addition, subtraction, and multiplication, i.e., the sum and product of any integers is an integer. Moreover, $\bbZ$ forms an unital ring. If we restrict to natural numbers, $\bbN$ is not closed under subtraction but closed under exponentiation. 
    Hence, linear algorithms on integers are limited in their operations as well.
    Feasible operations encompass
    \vspace*{-0.75\baselineskip}
    \unskip
    \begin{multicols}{2}\unskip
        \begin{itemize}\unskip
            \item $x+a$ with $a\in\bbZ$,
            \item $mx$ with $m\in\bbZ$, 
            \item $x \pmod{p}$ with $p\in\bbZ$,
            \item $x^b$ with $b\in\bbN$, 
            \item $c^x$ for $x\geq 0$ with $c\in\bbZ$,
        \unskip
        \end{itemize}\unskip
    \end{multicols}\unskip
    \vspace*{-0.75\baselineskip}
    and combinations thereof.
    In some dynamic systems governed by these mappings, it is possible to discover a linear representation based on a state dimension smaller than the linear complexity of the sequence it generates. 
    The result highlighted in Example~\ref{exmp:scalar} even demonstrates the possibility of scalar representations. 
    However, whether such a transformation is generally achievable remains an open question.
    This aligns with related conclusions in Koopman theory, where identifying the optimal set of observables remains an unsolved challenge.
    Often, this selection determines the accuracy of the linear representation, but here, it impacts the necessary lifting dimension.
    In particular, the chosen lifting functions in the examples above are related to the eigenfunctions of the Koopman operator. 
    Thus, if the eigenfunctions are known, the lifting functions can be chosen accordingly to obtain a lower dimensional linear representation of the system.
\end{rem}
We note that the operations above include also dynamical systems generating sequences which are not considered in the above examples. A (reduced) linear representation for systems generated, e.g., by the dynamics $x_{k+1} = (mx_k + a)^b$ or $x_{k+1} = mx_k^b + a$ or $x_{k+1} = c^{x_k}$ for arbitrary $m,a,c\in\bbZ$ is typically challenging to obtain and part of ongoing research.

\vspace*{-0.35\baselineskip}
\section{Data-based Koopman representation}\label{sec:data-Koopman}
\vspace*{-0.35\baselineskip}
In this section, our aim is to understand the behavior of the presented encryption methods using EDMD. 
To achieve this, we gather a set of sample data denoted as $\{x_k\}_{k=0}^N$ from the system~\eqref{eq:dynamical-system}. 
This data is then organized into matrices $X$, $X_+$, and lifted data matrices $Z$, $Z_+$ according to Section~\ref{sec:EDMD} and the $(q+1)$-dimensional lifted state $z=h(x)$. 
Due to the definition of $Z$, we immediately observe $\rank(Z) \leq \min(q+1,N)$.%
\begin{assum}\label{ass:number-of-data-samples}
    The data trajectory is of length $N\geq \tilde{q} + 1$, and the lifted state dimension satisfies $q\geq \tilde{q}$.
\end{assum}
We can now exploit the results in Theorem~\ref{thm:minimal-number-observables-DH} on the necessary lifting dimension $\tilde{q}+1$ for DH to characterize the rank of $Z$.
\begin{cor}\label{cor:EDMD-DH-full-row-rank-Z}
    The matrix $Z$ satisfies $\rank(Z) = \tilde{q}+1$ for DH cryptosystems.
\end{cor}%
\vspace*{-\baselineskip}
\begin{pf}%
    This is a direct application of Theorem~\ref{thm:minimal-number-observables-DH} and Corollary~\ref{cor:linear-system-for-q}. 
    First, since $z_{k+1} = z_{k-\tilde{q}} - z_{k-\tilde{q}+1} + z_{k}$, we deduce that every $z_j$ for $j\geq \tilde{q}+1$ is linearly dependent on $\{z_k\}_{k=0}^{\tilde{q}}$. 
    Thus, $Z$ can only satisfy $\rank(Z)\leq \tilde{q}+1$.
    Moreover, Theorem~\ref{thm:minimal-number-observables-DH} guarantees linear independence of $\{z_k\}_{k=0}^{\tilde{q}}$, such that we conclude $\rank(Z)= \tilde{q}+1$.
\end{pf} 
\vspace*{-\baselineskip}
A data-based linear representation of the underlying cryptosystem can then be obtained via EDMD.
In particular, we consider the finite-dimensional Koopman representation $z_{k+1} = \hat{A} z_k$, where $\hat{A}$ results via the least-squares problem $
    \hat{A} = \argmin_{A} \| Z_+ - A Z \|_\mathrm{F}
$.
This problem has a unique solution if $Z$ has full row rank. 
Thus, we restrict ourselves for the data-driven reconstruction to liftings based on $q=\tilde{q}$ leading to full row rank of $Z$ due to Corollary~\ref{cor:EDMD-DH-full-row-rank-Z}. 
Then, $\hat{A} = (Z_+Z^\top)(ZZ^\top)^{-1}$. We note that the solution satisfies $\hat{A}=A$ with $A$ defined as in Corollary~\ref{cor:linear-system-for-q}.
Hence, the discussion in Section~\ref{sec:reconstruction-of-secrets} about the reconstruction of secrets of the DH cryptosystem is also valid based on the data-driven system matrix $\hat{A}$.

Similarly, the next corollary characterizes the rank of $Z$ for RSA cryptosystems based on Theorem~\ref{thm:minimal-number-observables-RSA}. Note that we only obtain an upper bound for RSA since Theorem~\ref{thm:minimal-number-observables-RSA} also only yields an upper bound on the necessary lifting dimension.%
\begin{cor}%
    The matrix $Z$ satisfies $\rank(Z) \leq \tilde{q}+1$ for RSA cryptosystems.
\end{cor}%
\vspace*{-\baselineskip}
\begin{pf}%
    Here, we exploit Theorem~\ref{thm:minimal-number-observables-RSA}. 
    In particular, we use~\eqref{eq:periodicity-xk-RSA}, i.e., $z_{k+1} = z_{k-\tilde{q}}$, to deduce that every $z_j$ for $j\geq \tilde{q}+1$ is linearly dependent on $z_0,...,z_{\tilde{q}}$.
    Thus, $Z$ satisfies $\rank(Z)\leq \tilde{q}+1$.
\end{pf}%
\vspace*{-\baselineskip}
Since the data matrix $Z$ is not guaranteed to have full (row) rank in general, the least-squares estimate corresponding to EDMD might not have a (unique) solution. 
In particular, it is possible that there exists a smaller lifting dimension $\check{q}+1$ leading to a linear representation of the RSA cryptosystem. The smallest possible lifting dimension is characterized by $\check{q}\coloneqq \rank(Z)-1$. 
Hence, if the data matrix $Z$ has full row rank, the upper bound of Theorem~\ref{thm:minimal-number-observables-RSA} is indeed attained and $\check{q}=\tilde{q}$. 
Otherwise, a smaller lifting dimension is sufficient for the linear representation. 
This linear representation, i.e., the vector $\alpha$ of the state transition matrix $A$ in~\eqref{eq:Koopman-A-matrix-RNF}, is recovered via EDMD.
To be precise, we restrict ourselves to the first $\check{q}+1$ rows in $Z$ and $Z_+$ denoted by $\check{Z}$ and $\check{Z}_+$, respectively.
Then, $\check{A} = \argmin_A\|\check{Z}_+ - A \check{Z}\|_\mathrm{F}$ defines the corresponding data-based companion matrix of the lifted linear system with minimal dimension, where $\sum_{j=0}^{\check{q}}\alpha_j=1$. 
We note that this linear system of dimension $\check{q}+1$ contains all the information of the underlying RSA cryptosystem and can be used for the reconstruction of secrets according to Section~\ref{sec:reconstruction-of-secrets}.
A more rigorous, analytical investigation of the minimal lifting dimension depending on a given $p$ and $m$ is left for future work.

\vspace*{-0.35\baselineskip}
\section{Summary and Outlook}\label{sec:summary-outlook}
\vspace*{-0.35\baselineskip}
In this paper, we demonstrated how cryptosystems based on the DH key exchange or RSA can be redefined as nonlinear dynamical systems.
Relying on concepts from Koopman theory, we obtained a corresponding linear system for these nonlinear cryptosystems by lifting the state into a higher-dimensional space.
For a specific choice of observables, we provided the minimal required state dimension to precisely represent DH and also established a tight upper limit on the state dimension for a linear representation of RSA.
We linked this dimension to the notion of complexity, drawing parallels to the classical concept of linear complexity, and showcased how the Koopman approach can offer enhanced capabilities for analyzing certain discrete dynamics.
Further, we presented a purely data-driven analysis of the considered encryption schemes using EDMD. 
Building upon the linear system, we proposed solutions for decrypting the secrets within the cryptosystem.
Just like conventional number-theoretic methods, the proposed approach suffers from the sheer size of the numbers used in these cryptosystems, which renders the decryption computationally intractable in practice.
However, the derived formulas offer a novel and interesting perspective.
For example, an interesting question is whether the analysis of closed-loop systems based on encrypted data can benefit from the dynamical system representation of the cryptosystem itself.
\\
Further work in this area could focus on the extension to general finite fields, different cryptosystems, e.g., the Paillier cryptosystem~\citep{paillier:1999}, or digital signature schemes, e.g., the ElGamal signature scheme~\citep{elgamal:1985} or its more widely used variant called Digital Signature Algorithm which was developed at the National Security Agency~\citep{kerry:gallagher:2013}.
Another interesting research question is whether the herein gathered insights can be leveraged to possibly certify new schemes based on the complexity.
\vspace*{-0.75\baselineskip}

\bibliography{./literature.bib}

\vspace*{-2\baselineskip}
\appendix
\section{Reconstruction of $x_k$ based on observables on the complex unit circle}\label{sec:appendix:reconstruction-xk-complex-unit-circle}
\begin{figure*}[!t]
   \begin{equation}\label{eq:app-A0-structure}
   		\left[\begin{smallmatrix}
   			1
   			& m
   			& \cdots
   			& m^{\tilde{q}-1} + p\sum_{j=2}^{\tilde{q}-1} m^{\tilde{q}-1-j} \ell_j
   			& p-1
   			\\
   			m
   			& m^2 + p\ell_2
   			& \cdots
   			& p-1
   			& p-m
   			\\
   			\vdots 
   			& \vdots 
   			& \ddots 
   			& \vdots 
   			& \vdots 
   			\\
   			m^{\tilde{q}-1} + p\sum_{j=2}^{\tilde{q}-1} m^{\tilde{q}-1-j} \ell_j
   			& p-1  
   			& \cdots
   			& p-m^{\tilde{q}-2} -p\sum_{j=2}^{\tilde{q}-2} m^{\tilde{q}-2-j} \ell_j
   			& p-m^{\tilde{q}-1} -p\sum_{j=2}^{\tilde{q}-1} m^{\tilde{q}-1-j} \ell_j
   			\\
   			1 
   			& 1 
   			& \cdots 
   			& 1 
   			& 1
   		\end{smallmatrix}\right]
   	\end{equation}
   	\hrule
\end{figure*}
\vspace*{-\baselineskip}
We start with the representation $
    x_k \mathbb{1} = M^{-1}(\tilde{z}_k + \tilde{\ell} p)
$ and need to find suitable $\tilde{\ell}_j\in\bbZ$.
To this end, we parametrize $\tilde{\ell}_j$ as 
\begin{equation*}
    \tilde{\ell}_j = \alpha_{1,j} + \alpha_{2,j} - \alpha_{3,j} x_k + m^{j+1} \alpha_{4,j} + \alpha_{3,j}\alpha_{4,j}p
\end{equation*}
for some $\alpha_{1,j},\alpha_{2,j},\alpha_{3,j},\alpha_{4,j}\in\bbZ$.
Then, using $\tilde{\ell}_j$ in $m^{j+1} x_k = \tilde{z}_{k,j} + p\tilde{\ell}_j$ yields
\begin{multline*}
    (m^{j+1} + \alpha_{3,j} p) x_k 
    \\
    = (\tilde{z}_{k,j} + \alpha_{2,j} p) + \alpha_{1,j} p + \alpha_{4,j} p (m^{j+1} + \alpha_{3,j} p).
\end{multline*}
Consequently, solving for $x_k$ leads to
\begin{equation*}
    x_k = \frac{(\tilde{z}_{k,j} + \alpha_{2,j} p) + \alpha_{1,j} p}{m^{j+1} + \alpha_{3,j} p} + \alpha_{4,j} p.
\end{equation*}
Thus, choosing $\alpha_{2,j}$, $\alpha_{3,j}$, and $\alpha_{4,j}$ can be omitted by introducing several modulo operations. 
More explicitly, we only need to choose $\alpha_{1,j}$ such that
\begin{equation*}
    x_k = \frac{(\tilde{z}_{k,j}\pmod{p}) + \alpha_{1,j} p}{m^{j+1} \pmod{p}} \pmod{p}
\end{equation*}
is satisfied for $x_k\in\bbN_{1:p-1}$. 
Hence, the above procedure can reconstruct $x_k$ based on the lifted state $z_k$.

\vspace*{-0.6\baselineskip}
\section{Full rank of $(\hat{A}|\hat{b})$}\label{sec:appendix:full-rank-Ab}
\vspace*{-0.75\baselineskip}
We consider the composite matrix $(\hat{A}|\hat{b})$ corresponding to the system of equations~\eqref{eq:LGS-general-q-Euler-reduced-full-rank}.
In the following, we establish that $(\hat{A}|\hat{b})$ has full rank. 
Note that $x_k=m^k + \gamma_kp \in\bbN_{1:p-1}$ for $k\in\bbN_{0:p-2}$,
where $\gamma_0=\gamma_1=0$ and $\gamma_k\in\bbZ$ for $k\geq 2$.
Hence, 
\begin{equation*}
    x_{k}
    = \begin{cases}
        m^k
        & \text{for }k\in\bbN_{0:1},
        \\
        m^k + \gamma_k p 
        & \text{for }k\in\bbN_{2:\frac{p-1}{2}-1}.
    \end{cases}
\end{equation*}
Moreover, for $k\in\bbN_{2:\frac{p-1}{2}-1}$ we have 
\begin{align*}
    x_{k} &= m^{k} + \gamma_{k} p 
    = m(m^{k-1} + \gamma_{k-1} p) \pmod{p}
    \\
    &= m(m^{k-1} + \gamma_{k-1}p) + \ell_{k}p,
\end{align*}
i.e., $\ell_{k} = \gamma_{k} - m \gamma_{k-1}$. 
Thus, using that $\gamma_1=0$, we obtain
\begin{equation*}
    x_{k}
    = \begin{cases}
        m^k
        & \text{for }k\in\bbN_{0:1},
        \\
        m^k + p \sum_{j=2}^{k} m^{k-j} \ell_j
        & \text{for }k\in\bbN_{2:\frac{p-1}{2}-1}.
    \end{cases}
\end{equation*}
Hence, we conclude the particular structure of $(\hat{A}|\hat{b})$ given in~\eqref{eq:app-A0-structure} for $\tilde{q}=(p-1)/2$.
More precisely, this structure ensures linear independence of the columns of $(\hat{A}|\hat{b})$ and, hence, full rank of $(\hat{A}|\hat{b})$. 
In particular, this can be observed due to the last row of ones, the fact that $m$ is a generator of $\bbZ_p^*$, and that the modulo operator is indeed a nonlinear operation, i.e., there exists a $j\in\bbN_{2:\frac{p-1}{2}-1}$ for which $\ell_j\neq 0$.

\vspace*{-0.6\baselineskip}
\section{Proof of Lemma~\ref{lm:Euler-criterion-generalization}}\label{sec:appendix:Euler-criterion-generalization-proof}
\vspace*{-0.35\baselineskip}
Note that condition~\eqref{eq:Euler-criterion-generalization} reduces to Euler's criterion for $p_2 = 1$ as then $p=p_1p_2=p_1$ is an odd prime number and $\varPhi(p)=p-1$. Thus, we only show the case $p_2\neq 1$ where both $p_1$ and $p_2$ are distinct odd prime numbers resulting in $m^{\varPhi(p)/2}\pmod{p}=1$. 
\\
Consider two distinct odd prime numbers $p_1,p_2$, i.e., $\gcd(p_1,p_2)=1$. Then, Bézout's identity~\citep{bezout:1779} yields the existence of two integers $\alpha_1,\alpha_2$ such that 
\begin{equation*}
    \alpha_1 p_1 + \alpha_2 p_2 = \gcd(p_1,p_2) = 1.
\end{equation*}
Thus, 
\begin{alignat*}{2}
    1 &\overset{{\pmod{p_1}}}{=} \alpha_1 p_1 + \alpha_2 p_2 & &\overset{{\pmod{p_1}}}{=} \alpha_2 p_2, \\
    1 &\overset{{\pmod{p_2}}}{=} \alpha_1 p_1 + \alpha_2 p_2 & &\overset{{\pmod{p_2}}}{=} \alpha_1 p_1.
\end{alignat*}
Let $p=p_1p_2$ and $\xi_1 = m^{\varPhi(p)/2} \pmod{p_1}$ and $\xi_2 = m^{\varPhi(p)/2} \pmod{p_2}$. In the following, we use 
\begin{equation}\label{eq:proof:phi-p-factorization}
    m^{\frac{\varPhi(p)}{2}} = m^{\frac{(p_1-1)(p_2-1)}{2}}
\end{equation}
and
\begin{equation}\label{eq:proof:modulo-exponent}
    a^k \pmod{b} = (a \pmod{b})^k \pmod{b}
\end{equation}
for $a,b,k\in\bbN$. Then,
\vspace*{-2.25\baselineskip}
\small
\begin{alignat*}{2}
    \xi_1 
    &\overset{\eqref{eq:proof:phi-p-factorization}}{=} 
    m^{\frac{(p_1-1)(p_2-1)}{2}} \pmod{p_1} 
    & &\overset{\eqref{eq:proof:modulo-exponent}}{=} 
    \left[
        m^{\frac{p_1-1}{2}} \pmod{p_1}
    \right]^{p_2 - 1} \\[-0.5ex]
    &\overset{\eqref{eq:Euler-criterion}}{=} 
    \left[
        \pm 1 
    \right]^{p_2 - 1}
    = 1,
    \\
    \xi_2 
    &\overset{\eqref{eq:proof:phi-p-factorization}}{=} 
    m^{\frac{(p_1-1)(p_2-1)}{2}} \pmod{p_2}
    & &\overset{\eqref{eq:proof:modulo-exponent}}{=} 
    \left[
        m^{\frac{p_2-1}{2}} \pmod{p_2}
    \right]^{p_1 - 1} \\[-0.5ex]
    &\overset{\eqref{eq:Euler-criterion}}{=} 
    \left[
        \pm 1
    \right]^{p_1 - 1}
    = 1,
\\[-1.5\baselineskip]
\end{alignat*}
\normalsize
where we use in the last equation that $p_1-1$ and $p_2-1$ are even as $p_1,p_2$ are odd prime numbers. By using the Chinese Remainder Theorem~\citep{gauss:1966} we get 
\vspace*{-0.8\baselineskip}
\begin{align*}
    m^{\frac{\varPhi(p)}{2}} \pmod{p_1p_2} 
    &= \xi_2 \alpha_1 p_1 + \xi_1 \alpha_2 p_2 \pmod{p_1p_2} \\
    &= \alpha_1 p_1 + \alpha_2 p_2 \pmod{p_1p_2} \\
    &= 1,
\\[-1.5\baselineskip]
\end{align*}
where we exploit $\xi_1=\xi_2=1$ for the second equation and Bézout's identity for the last equation. Thus, we obtain $m^{\varPhi(p)/2}\pmod{p} = 1$ for $p_1,p_2$ being distinct odd prime numbers. 
\null\hfill$\square$

\end{document}